\documentclass[10pt]{article}

\usepackage{fullpage}
\usepackage{times}
\usepackage{graphicx}
\usepackage{psfrag}
\usepackage{amsmath,amsthm}
\usepackage{amsfonts}
\usepackage{color}
\usepackage{authblk}
\usepackage{float}
\floatstyle{plaintop}
\restylefloat{table}

\newcommand{\var}{\operatorname{Var} }
\usepackage{authblk}

\makeatletter
\renewcommand{\@biblabel}[1]{\quad#1.}
\makeatother

\usepackage[labelfont=bf,labelsep=period,justification=raggedright]{caption}

\begin{document}

\bibliographystyle{plos2009}

\title{Reconstructing Native American Migrations from Whole-genome and Whole-exome Data}

\author[1,2$\ast$]{Simon Gravel}
\author[3,$\dagger$]{Fouad Zakharia}
\author[3,$\dagger$]{Andres Moreno-Estrada}
\author[3,4,$\dagger$]{Jake K. Byrnes}
\author[3,5]{Marina Muzzio}
\author[6]{ Juan L. Rodriguez-Flores}
\author[3,7]{Eimear E. Kenny}
\author[8]{Christopher R.  Gignoux}
\author[3]{Brian K. Maples}
\author[9]{Wilfried Guiblet} 
\author[10]{Julie Dutil}
\author[8,11]{Marc Via}
\author[3]{Karla Sandoval} 
\author[12]{Gabriel Bedoya} 
\author[13]{The 1000 Genomes project}
\author[9]{Taras K Oleksyk}
\author[14]{Andres Ruiz-Linares}
\author[8]{Esteban G. Burchard} 
\author[9]{Juan Carlos Martinez-Cruzado}
\author[3]{Carlos D. Bustamante}

\affil[1]{Department of Human Genetics, McGill University, Montr\'eal, QC, Canada}
\affil[2]{McGill University and G\'enome Qu\'ebec Innovation Centre, Montr\'eal, QC, Canada} 
\affil[3]{Department of Genetics, Stanford University, Stanford, CA, USA } 
\affil[4]{Ancestry.com DNA LLC, San Francisco, CA, USA}
\affil[5]{Laboratorio de Gen\'etica Molecular Poblacional, Instituto Multidisciplinario de Biolog\'ia Celular (IMBICE). CCT- CONICET-La Plata, Argentina and  Facultad de Ciencias Naturales y Museo, Universidad Nacional de La Plata, Argentina}
\affil[6]{Weill Cornell Medical College, New York, NY, USA}
\affil[7]{ Department of Genetics and Genomic Sciences, The Charles Bronfman Institute for Personalized Medicine, Center for Statistical Genetics, and Institute for Genomics and Multiscale Biology, Icahn School of Medicine at Mount Sinai, New York, NY USA}
\affil[8]{Department of Bioengineering and Therapeutic Sciences and Medicine, UCSF, San Francisco, CA, USA}  
\affil[9]{Department of Biology, University of Puerto Rico at Mayaguez, Puerto Rico}
\affil[10]{Department of Biochemistry, Ponce School of Medicine and Health Sciences, Ponce, PR}
\affil[11]{Department of Psychiatry and Clinical Psychobiology, University of Barcelona}
\affil[12]{Universidad de Antioquia, Medell\'in, Colombia}
\affil[13]{A list of contributors can be found at http://www.1000genomes.org/participants}
\affil[14]{Department of Genetics, Evolution and Environment, University College London, London, UK}
\affil[$\dagger$]{These authors contributed equally} 
\affil[$\ast$]{ E-mail: simon.gravel@mcgill.ca}

\maketitle

\begin{abstract}

There is great scientific and popular interest in understanding the genetic history of populations in the Americas.   We wish to understand when different regions of the continent were inhabited, where settlers came from, and how current inhabitants relate genetically to earlier populations.  Recent studies unraveled parts of the genetic history of the continent using genotyping arrays and uniparental markers.  The 1000 Genomes Project provides a unique opportunity for improving our understanding of population genetic history by providing over a hundred sequenced low coverage genomes and exomes from Colombian (CLM), Mexican-American (MXL), and Puerto Rican (PUR) populations. Here, we explore the genomic contributions of African, European, and especially Native American ancestry to these populations. Estimated Native American ancestry is $48\%$ in MXL, $25\%$ in CLM, and $13\%$ in PUR. Native American ancestry in PUR is most closely related to populations surrounding the Orinoco River basin, confirming the Southern America ancestry of the Ta\'ino people of the Caribbean. We present new methods to estimate the allele frequencies in the Native American fraction of the populations, and model their distribution using a demographic model for three ancestral Native American populations. These ancestral populations likely split in close succession: the most likely scenario, based on a peopling of the Americas $16$ thousand years ago (kya), supports that the MXL Ancestors split $12.2$kya, with a subsequent split of the ancestors to CLM and PUR $11.7$kya. The model also features effective populations of $62,000$ in Mexico, $8,700$ in Colombia, and $1,900$ in Puerto Rico. Modeling Identity-by-descent (IBD) and ancestry tract length, we show that post-contact populations also differ markedly in their effective sizes and migration patterns, with Puerto Rico showing the smallest effective size and the earlier migration from Europe. Finally, we compare IBD and ancestry assignments to find evidence for relatedness among European founders to the three populations.

\end{abstract}

\section*{Author summary}

Populations of the Americas have a rich and heterogeneous genetic and cultural heritage that draws from a diversity of pre-Columbian Native American, European, and African populations. Characterizing this diversity facilitates the development of medical genetics research in diverse populations and the transfer of medical knowledge across populations.  It also represents an opportunity to better understand the peopling of the Americas, from the crossing of Beringia to the post-Columbian era. Here we take advantage sequencing of individuals of Colombian (CLM), Mexican (MXL), and Puerto Rican (PUR) origin by the 1000 Genomes project to improve our demographic models for the peopling of the Americas. 

The divergence among African, European, and Native American ancestors to these populations enables us to infer the continent of origin at each locus in the sampled genomes.  The resulting patterns of ancestry suggest complex post-Columbian migration histories, starting later in CLM  than in MXL and PUR.  

Whereas European ancestral segments show evidence of relatedness, a demographic model of synonymous variation suggests that the Native American Ancestors to MXL, PUR, and CLM panels split within a few hundred years over 12 thousand years ago. Together with early archeological sites in South America, these result support rapid divergence during the initial peopling of the Americas.




\section*{Introduction}

The 1000 Genomes project \cite{GenomesProjectConsortium:2012co}  released sequence data for 66 Mexican-American (MXL), 60 Colombian (CLM), and 55 Puerto Rican (PUR) individuals using an array of technologies  including low-coverage whole genome sequence data, high-coverage exome capture data, and OMNI 2.5 genotyping data.  These data  provide a unique window into the settlement of the Americas that  complement archeological and the more limited genetic data previously available. Here we interpret these data to answer basic questions about the pre- and post-Columbian demographic history of the Americas. 

People reached the Americas by crossing Beringia during the Last Glacial Maximum, likely between 16-20 kya (see e.g. \cite{Orourke:2010jq, LuisLanata:2008cf,Goebel:2008vn,Dillehay:2009dx}).  The presence of early South American sites such as Monte Verde \cite{Dillehay2008} suggests a rapid occupation of the continent, which is supported also by recent mitochondrial DNA studies \cite{Bodner:2012ew}. A coastal route has been proposed to explain this rapid expansion (e.g., \cite{Hurst:1943wj,  Dillehay2008, Bodner:2012ew}), but other migration routes, possibly concurrent, have also been proposed (see. e.g., \cite{Dillehay:2009dx, Meltzer:2009wq}, and references therein). This original peopling of the Americas, followed by European contact starting in 1492 and substantial African slave trade starting in 1502, have created a diverse genetic heritage in American populations.  

The initial settlement of the Caribbean has been much debated (e.g. \cite{rouse1992tainos,Ramos:2010vd,veloz1991panorama} and references therein). People reached the islands around 7 kya, probably from a Mesoamerican source \cite{Ramos:2010vd}. Around 4.5 kya, a second wave of migrants probably reached the islands, likely coming from the Orinoco Delta or the Guianas in South America and speaking Arawakan languages  (see \cite{ Hopper:2008uo} and references therein).   By approximately 1.3 kya, they had established large Ta\'ino communities through the Greater Antilles, including Puerto Rico.

The earliest available account reports 600,000 Native Americans in Puerto Rico at the time of European arrival, not counting women and children (V\'azquez de Espinosa 1629). More conservative estimates suggest 110,000 individuals \cite{moscoso2008caciques}, and as few as 30,000 inhabitants in 1508 \cite{alegria1999historia}. All references agree that the Native American population was subsequently largely decimated through disease, forced labor, emigration, and war. 
Despite the bottleneck at contact, admixture and the subsequent population growth on the Island resulted in a Native American genetic contribution averaging $15.2\%$ of the modern population of $3.77$ million \cite{via:2011hs}. 

The MXL were sampled in Los Angeles, USA and the CLM in Medellin, Colombia. These panels represent urban populations, but recent urbanization means that they derive ancestry from larger geographic areas. Among respondents to the 2005 Colombia Census in Medellin, $61.3\%$ were born in the city, and  $38\%$ were born in another part of Colombia, with a sizable proportion from the surrounding  Department of Antioquia. Given this high rate of within-country migration, but a relatively low rate of migration from outside Colombia, we can think of the sample as representing a diverse sample from Antioquia. Similarly, the 1.2M Angelenos of Mexican origin in the 2010 US census represent the added contributions of multiple waves of migrations starting with the city's foundation in 1781 and received contributions from diverse states.

The use of genetic data to study Native American history is well established. The bulk of these studies rely on Y chromosome  \cite{Underhill:1996wq,Lell:1997uy,Bianchi:1998jq,Karafet:1999bp,  Bortolini:2003bk,Mesa:2000ht,Bortolini:2002ci,Bailliet:2009jn} and mitochondria DNA (mtDNA)  \cite{Torroni:1993ur, Achilli:2008ba,Kumar:2011iy, Malhi:2010fm,Perego:2009bk,Perego:2010hu, Tamm:2007bn,Mesa:2000ht, Sandoval:2009el,Bodner:2012ew, Bonatto2009AS, Mulligan:2008dd, Fagundes:2008jn}, with a number of studies using increasingly dense sets of autosomal markers  \cite{Mesa:2000ht, Wang:2007be, Rojas:2010ii, Yang:2010ip,  Scliar:2012hw, Reich:2012cs}. Such studies provided evidence for a bottleneck recovery into the Americas 16-12 kya (e.g., \cite{Mulligan:2008dd,Fagundes:2008jn}), and for complex models of migrations and admixture within Native groups \cite{Reich:2012cs}. 

In this article, we use the 1000 Genomes data and a diversity of  population genetic tools to delve deeper in the founding of the Puerto Rican, Mexican, and Colombian populations. 
To propose models for Native American demography, we must first quantify the African, European, and Native American contributions to these populations. Because of strong sex-asymmetric migrations, autosomal and sex-linked markers exhibit substantial differences in ancestry proportions \cite{MartinezCortes:2012ed, Rubicastellanos:2009df, Bedoya:2006ei, Martinez:2005RT,Bolnick:2006fl, CarvajalCarmona:2000bp}. Focusing on the autosomal regions, we infer the locus-specific pre-Columbian continental ancestry in each sample, and estimate the timing and intensity of different migration waves that contributed to these populations. Using identity-by-descent analysis, we identify relatedness among the different ancestral groups and estimate recent effective population sizes.

We also propose a three-population model based on the diffusion approximation to study the distribution of allele frequencies across the Native American ancestors of the MXL, PUR, and CLM. We present statistical methods that take advantage of admixture linkage patterns to disentangle the histories of each continental group. The large sample of sequence data allows for the joint inference of split times and effective population sizes among the Native ancestors to the three panels.  Finally, through an expectation maximization (EM) framework, we estimate genome-wide allele frequencies in the inferred Native components of MXL, CLM, and PUR genomes.

\begin{figure}
\scalebox{0.6}{\includegraphics{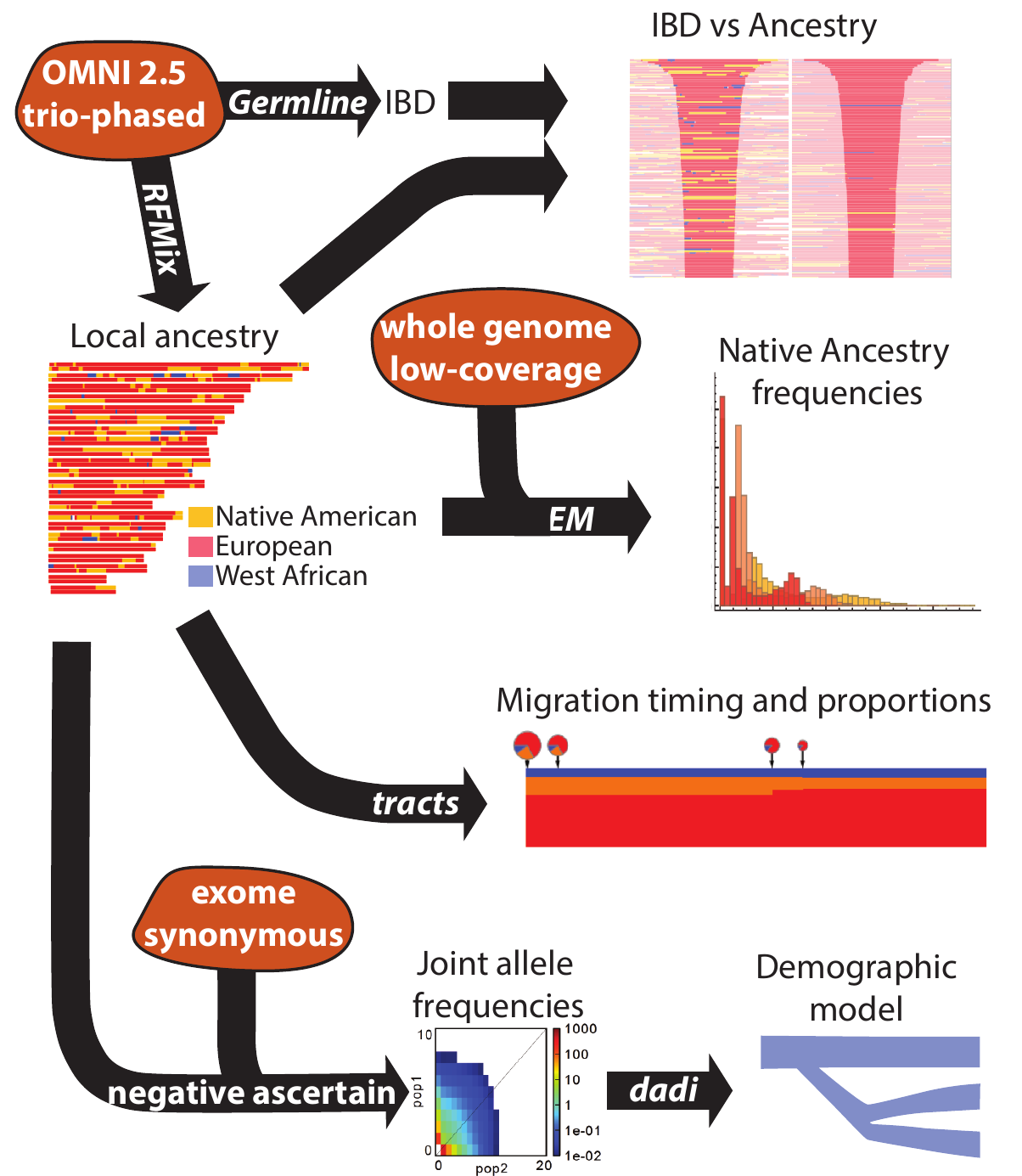}}
\caption{Schematic of the data and analysis pipelines used in this article. The three types of 1000 Genomes data are shown in orange: whole-genome, low-coverage data; exome capture; and genotyping chip.  Only genotyping chip data was available in trio-phased form; for the other two datasets we used unphased genotypes. \label{flow} Among the analysis approaches (black arrows), the EM and the negative ascertainment analysis are novel: they are presented in the Methods section. }
\end{figure}

A broad summary of the data and analysis pipelines used in this article are displayed in Figure \ref{flow}.

\section*{Results}
\subsection*{Global ancestry proportions and clustering}

\begin{figure}
\scalebox{0.7}{\includegraphics[]{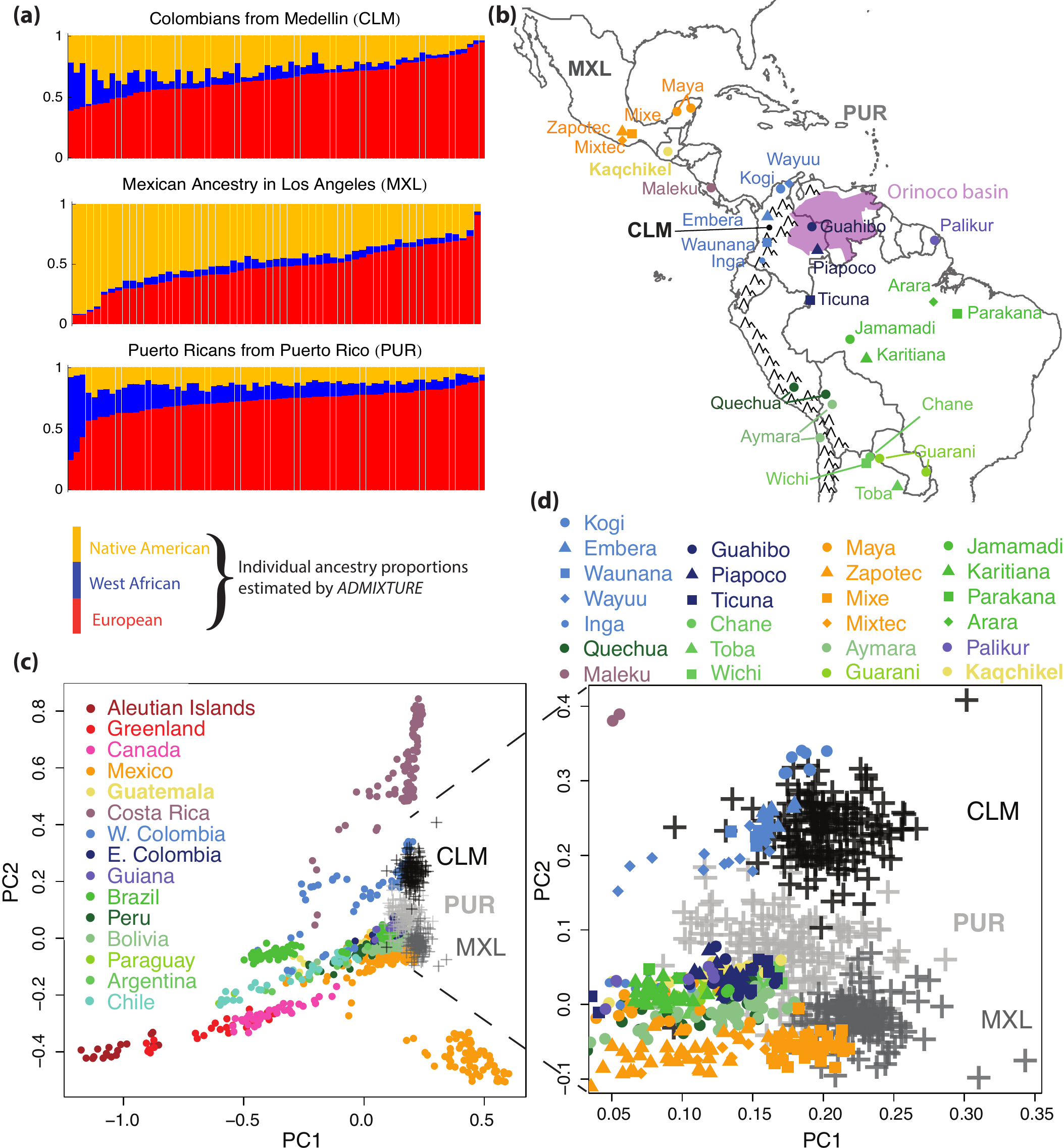}}

\caption{\label{PCA} (a) Individual ancestry proportions in the 1000 Genomes CLM, MXL, and PUR populations according to {\sc Admixture}, (b) Map showing the sampling locations for the populations most closely related to the Native components of the 1000 Genomes populations. (c) Principal component analysis restricted to genomic segments inferred to be of Native Ancestry in these populations, compared to a reference panel of Native American groups from \cite{Reich:2012cs}, pooled according to country of origin as a proxy for geography. Populations sampled across many locations are labeled according to the country of the centroid of locations. (d) Zoomed version of the PCA plot, showing specific Native American population labels, colored according to country of origin. }
\end{figure}

To estimate the global proportions of African, European, and Native American ancestry in the CLM, MXL, and PUR, we combined them with YRI, CEU, and a panel of Native American samples  \cite{Reich:2012cs}  and performed an {\sc admixture}  \cite{Alexander:2009bz} analysis (Figure \ref{PCA}(a)) and principal component analysis (Figure \ref{regPCA}). 

Dense genotyping arrays allow for inference of ancestry at the level of individual loci, using software such as   {\sc RFMix} \cite{Maples:2013fia}. Trio-phased OMNI data was used to generate such locus-specific ancestry calls for 66 CLM, 68 MXL, and 64 PUR individuals, including all sequenced individuals, as part of the 1000 Genomes Project.  Summing up the local ancestry contribution inferred by {\sc RFMix} provides an alternate estimate of ancestry proportions.

Using {\sc Admixture}, we find Native American proportions being $12.8\%$ 
in PUR, $25.6\%$ 
 in CLM, and $47.6\%$ 
 in MXL (Figure \ref{PCA}a). {\sc RFMix} finds values falling within $0.5$ percentage points of these values, and within one percentage point of the values inferred in the 1000 Genomes project through related methods \cite{GenomesProjectConsortium:2012co}. Estimates of African ancestry showed a larger difference across methods, with {\sc Admixture} ({\sc RFMix}) estimates at $14.8\% (11.7\%)$ in PUR,   $8.9\% (7.8\%)$ in CLM, and $5.4\% (4.2\%)$ in MXL.  

{\color{black}  The inferred Native American ancestry proportions are in good agreement with results from the GALA study \cite{Galanter:2012ig}, which reported proportions of  $12.4\%$ in Puerto Rico and $49.6\%$ in Mexico. The PUR result is also comparable to the $15.2\%$ of Native ancestry inferred in a different Puerto Rican sample \cite{via:2011hs}.  By contrast, none of the populations from Colombia in \cite{Rojas:2010ii} show median ancestry proportions quite similar to the CLM sample from Medellin, the closest being the sample from the surrounding Department of Antioquia, with $39\%$ Native, $6\%$ African and $52\%$ European.  
}

Figure \ref{PCA}(c-d) shows a principal component analysis restricted to segments of inferred Native ancestry \cite{MorenoEstrada:2013wx}.  We find that the MXL individuals cluster primarily with southern Mexican Native groups (mostly Mixe), and the CLM cluster primarily with the Embera, Kogii, and Wayu, all of which were sampled in Colombia North-West of the Andes, where Medellin is also located. The PUR clusters principally with populations South-East of the Andes, surrounding the Guyanas and the Orinoco River basin  (Ticuna, Guahibo, Palikur, Jamamadi, Piapoco), although a few populations from further south are also close in PCA space, particularly the Guaran\'i and the Chan\'e, together with some Kaqchikel, Toba,  and Wichi individuals. The Piapoco and the Palikur speak Arawakan languages. The other groups with known Arawakan-speaking ancestors in our panel are the Chan\'e, whose ancestors spoke Arawakan and likely originated in Guiana \cite{Moseley:2004wc}, and the Guarani, through gene flow from the Chan\'e \cite{Combes:2006gv}. Taken together, these clustering patterns support a demic diffusion of the Arawakan/Ta\'inos into Puerto Rico from a southern American route, and reduced gene flow between Native Americans groups living in the Andes or to the west, and groups living east of the Andes.

\subsection*{Ancestry tracts analysis}
\begin{figure}
\scalebox{0.3}{\includegraphics{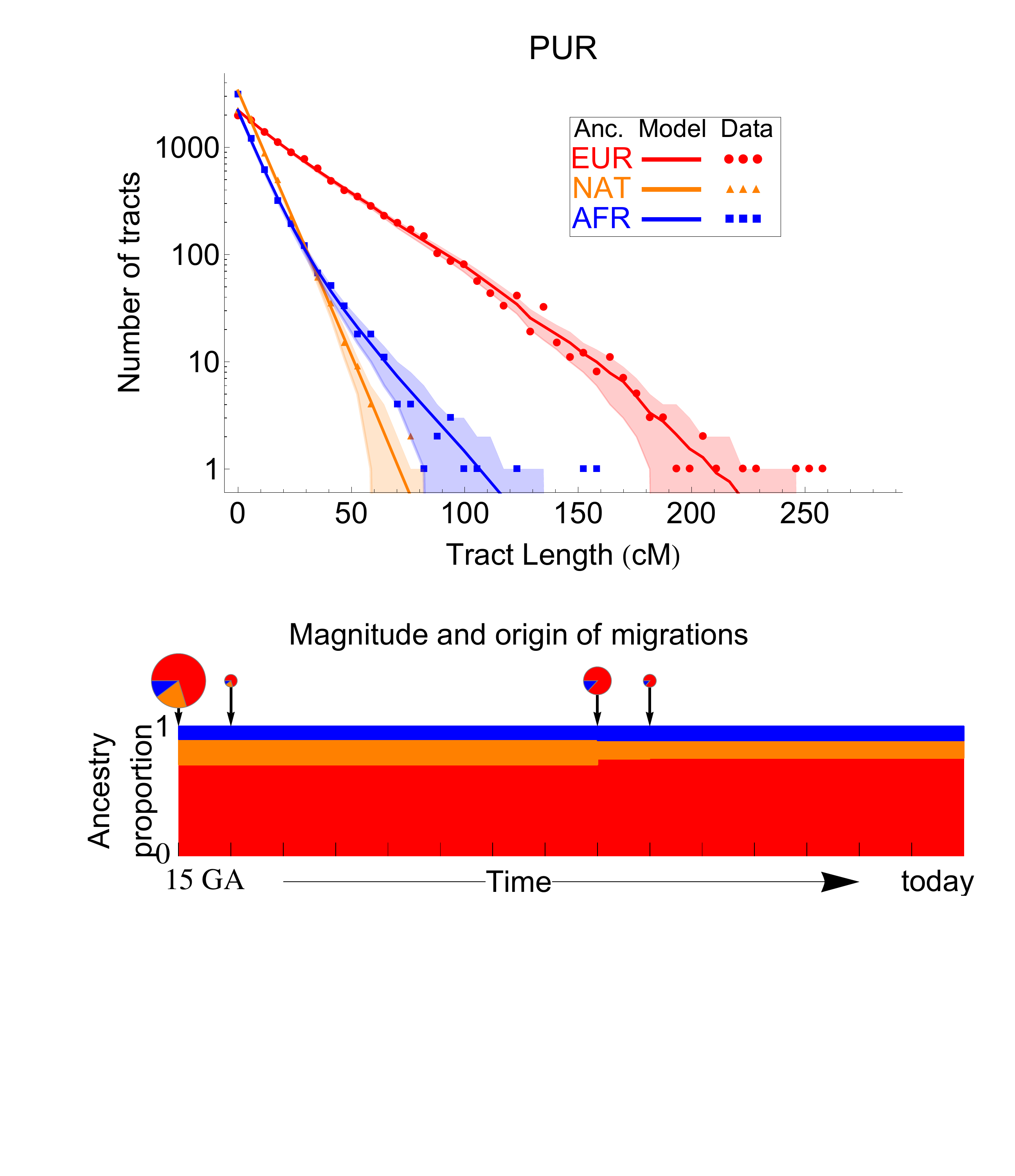}}
\scalebox{0.3}{\includegraphics{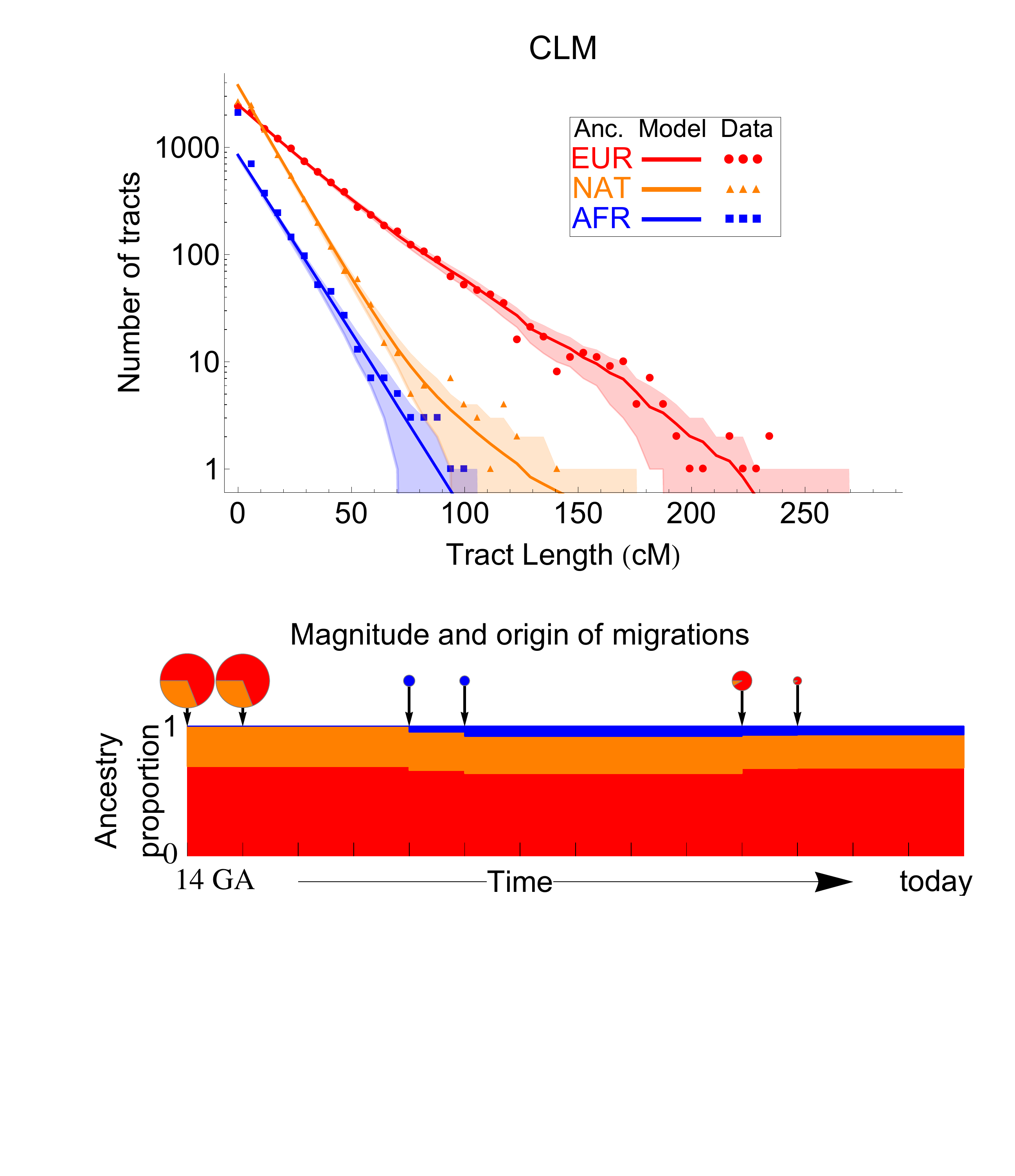}}
\caption{\label{PURtracts} Ancestry tract length distribution in PUR (a) and CLM (b) compared to the predictions of the best-fitting migration model. Solid lines represent model predictions and shaded areas are one standard deviation confidence regions surrounding the predictions, assuming a Poisson distribution of counts per bin. The best-fitting models are displayed under each graph. {\color{black} Pie charts sizes indicate the proportion of migrants at each generation, and the pie parts represent the fraction of migrants of each origin at a given generation. Migrants are taken to have uniform continental ancestry. `Single-pulse' admixture events occurring at non integer time in generations are distributed among neighboring generations: in the CLM, the inferred onset was 13.02 generations ago (ga). The model involves founding 14 ga, but almost complete replacement 13 ga. } At 30 years per generation \cite{Tremblay:2000jd}, 14.9 ga corresponds to $c.1566$, and 13 to $c. 1623$. Model parameters and confidence intervals are displayed in Table S1 in the Text S1 file.}
\end{figure}

{\color{black} Because continuous tracts of local ancestry are progressively broken down by recombination, the length distribution of continuous ancestry tracts can reveal details of the timing and mode of the migration processes. 
We used RFMix to infer ancestry tracts (Text \ref{suppmet}), and the software {\sc tracts} \cite{Gravel:2012ip}  to infer the migration rates and model likelihoods under different scenarios.  {\sc Tracts} can predict the distribution of ancestry block length for arbitrary models of time-varying migration, under the assumptions that the migrants are themselves not admixed, and that the admixed population follows Wright-Fisher reproduction. Since admixture only begins after two populations are in contact, the admixed population is founded when the second population arrives.  {\sc Tracts} determines the time and ancestry proportions at the onset of admixture and the time and magnitude of subsequent migrations by maximum likelihood.  Because of limited statistical power,  we start with a simple model in which each population contributes a single pulse of migration. We then progressively introduce models with additional periods of migration when justified by information criteria, as described in Text \ref{suppmet}. The models that best describe the data are shown in Figures \ref{PURtracts} and \ref{MXLtracts}. Parameters for these, together with confidence intervals obtained through bootstrap over individuals, are provided in Table S1 in the Text S1 file}.

For MXL, we considered a model introduced in \cite{Kidd:2012fi}: three populations start contributing migrants at the same time, but Europeans and Native Americans keep contributing at a constant rate. The best-fitting model has an onset of admixture 15.1 generations ago (ga), with a $95\%$ CI of $(13.7-17.1)$, in good agreement with \cite{Kidd:2012fi} despite a different genotyping chip and local ancestry inference method.  

In PUR, we found evidence for two periods of European and African migration, the first $14.9$ ga ($95\%$ CI $14.2-15.9$) and the most recent period at $6.8$ga ($95\%$ CI 5.9-8.8). This model is in excellent agreement with historical records, which suggest that isolated Native populations contributed little gene flow to the colony after the initial contact period, and that substantial slave trade and European immigration continued until the second half of the 19th century. We do not mean to imply that migrations actually occurred in exactly two distinct pulses-we do not have the resolution to distinguish more than two pulses per population. However, the inference of a migration pulse 6.8 ga indicates that migrations occurred during a period spanning this date.  This complex scenario, with multiple waves of migration from African and European individuals, is consistent with the observation that European and African ancestries vary across the island, whereas no evidence of such variation was found in Native ancestry \cite{via:2011hs}.   

The inferred onset of admixture in CLM is 13.0 ga ($95\%$ CI $12.5-13.9$), significantly later than that in both MXL and PUR and consistent with later European settlement in western Colombia compared to Mexico and Puerto Rico.  We also find evidence for a small but statistically significant second wave of Native American migration, 4.8 ga ($95\%$ CI 4-6). As above, this does not necessarily indicate a single, punctual event, but probable contact between an admixed population and Native American individuals during that period. By contrast, we find no evidence for continuing African gene flow in CLM.

\subsection*{Identity by descent analysis}   
\label{ibdresults}
 We used {\sc germline} \cite{Gusev:2009wp} and the trio-phased OMNI data above to identify segments identical-by-descent (IBD) within and across populations (see Text \ref{suppmet}). Not surprisingly, we found more IBD segments within populations (23936) compared to across populations (1440), and within-population segments were longer (Figure \ref{histIBD}). 

\begin{figure}
\scalebox{0.75}{\includegraphics{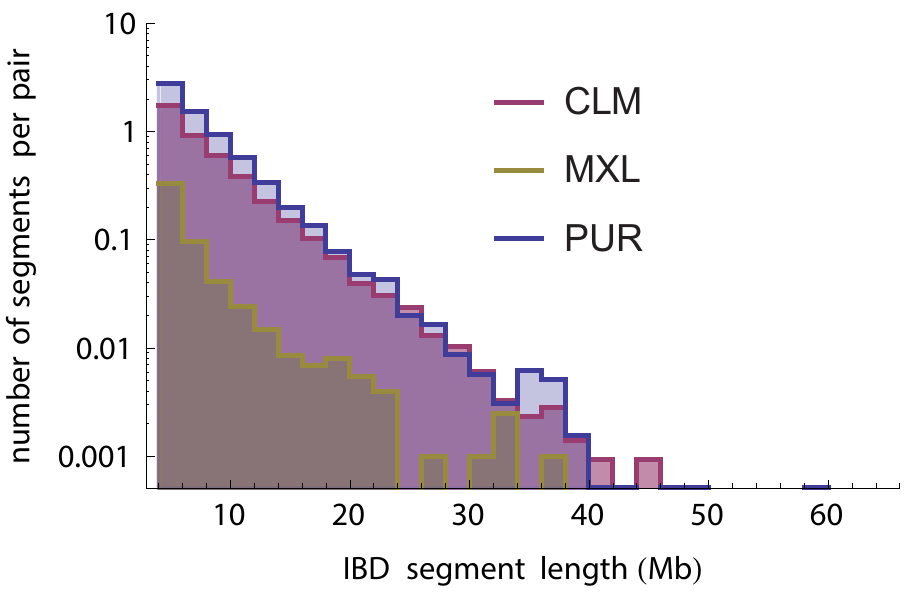}}
\caption{\label{IBDbypop} Number of IBD tracts by length bin in the three panel populations (independent of ancestry estimations), normalized by the number of individual pairs. The lower level of IBD in the MXL population indicate a much larger effective population size.}
\end{figure}

The MXL population exhibits significantly less within-population IBD compared to the other two panels (Figure \ref{IBDbypop}).
The amount of IBD among unrelated individuals can be used to infer the underlying population size under panmictic assumption: the larger a population, the more distant the expected relationship between any two individuals \cite{Palamara:2012LD}. Using IBD segments longer than 4cM, we infer effective population sizes of 140,000 in MXL, 15,000 in CLM, and 10,000 in PUR. As we will show, these largely reflect post-admixture population sizes.

We expect long IBD segments to be inherited from a recent common ancestor, and therefore to have identical continental ancestry. Comparing the {\sc RFMix} ancestry assignments on chromosomes that have been identified as IBD by {\sc germline} thus provides a measure of the consistency of the two methods (see \cite{Baran:2012ku}  for a related metric).  Rates of IBD-Ancestry mismatch   ranged from $2.6\%$ in segments of $5\mathrm{Mb}$ to less than $0.2\%$ for segments longer than 40Mb (Figure \ref{IILANC}). 

\begin{figure}
\scalebox{0.4}{\includegraphics{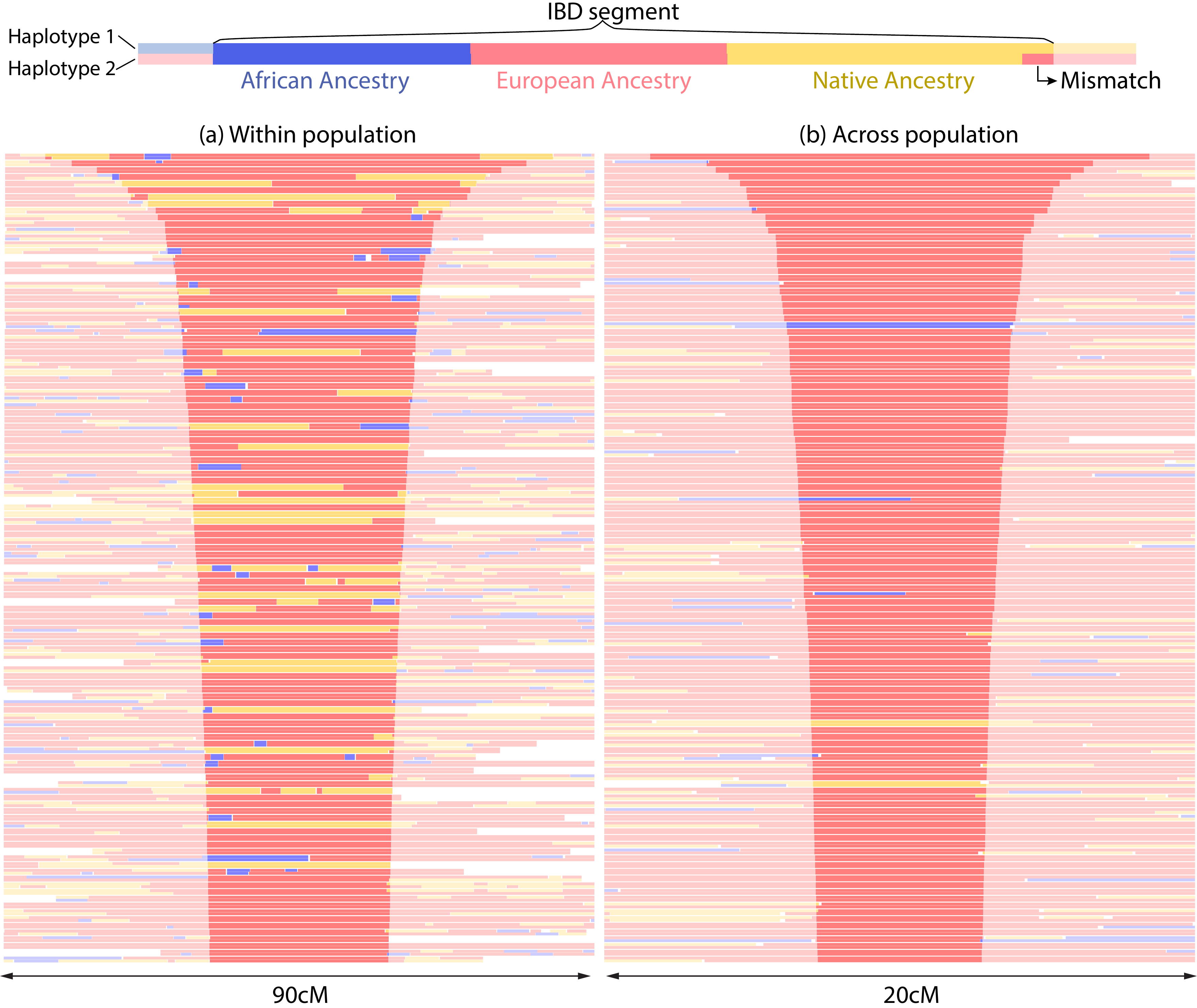}}
\caption{\label{IBDanc} (a) Local ancestry assignments in the neighborhood of the 120 longest inferred IBD segments within a population, (b) Local ancestry assignments in the neighborhood of the 120 longest inferred IBD segments across populations. Within inferred IBD segments, ancestry mismatches correspond $0.3\%$ error rate within population, and $0.5\%$ error rate across population.}
\end{figure}

Patterns of ancestry in IBD segments within a population differ markedly from those across populations (Figure \ref{IBDanc}):  IBD segments within populations contain many ancestry switches. 
This indicates that many common ancestors lived after contact, and that the effective population sizes estimated using IBD largely reflects post-contact demography. The IBD patterns in cross-population IBD segments exhibited fewer ancestry switches than a random control (Figure \ref{altcontrol}), as may be expected if common ancestors often predate the onset of admixture. Cross-population IBD segments were also found to be overwhelmingly of European origin: among the 120 longest cross-population IBD segments, 117 are in European-inferred segments, two are among Native segments, and one is among African segments. This is not due to overall ancestry proportions, as can be observed by considering the alternate (non-IBD) haplotypes at the same positions (Figure \ref{altcontrol}). This is likely a result of the colonization history, in which European colonists rapidly spread from a relatively specific region over a large continent. This interpretation is supported by the {\sc admixture} analysis (Figure \ref{Admixture312}), showing a common cluster of ancestry for the European component dominant in PUR, CLM, MXL, and Andean populations, but not in CEU, Eskimo-Aleut, and Na-Dene. 
Finally, we were interested in testing whether the relationship between IBD and ancestry can be used to date recombination events. The ancestry within an IBD segment represents the ancestry state of the most recent common ancestor. The shorter the IBD segment, the older the ancestor, and the less time available since the onset of admixture to create ancestry switch points through recombination.  
Indeed, we find that the \emph{density} of ancestry switch-points on IBD tracts increases with IBD tract length in PUR (bootstrap $p<0.001$, see Text \ref{suppmet}) and in MXL (bootstrap $p=0.03$), whereas the results are not significant in CLM. Thus we can use ancestry patterns in admixed populations not only to recognize recombination events but also to help date most recent common ancestors and recombination events (see Text \ref{suppmet} for details). The small amount of cross-population IBD among Native American tracts tells us that the ancestral Native populations were not as closely related as European founders, consistent with historical and anthropological data.

\subsection*{Demographic inference from sequence data}

 To infer split times and population sizes of the Native ancestors, we consider the joint site frequency spectrum (SFS). The SFS is informative of demography because stochastic differences in allele frequencies accumulate over time and at a rate that depends on population sizes. We use the diffusion-approximation framework implemented in $\partial a\partial i$ \cite{Gutenkunst:2009gs} to perform the inference.
We focus on synonymous sites in the 1000 Genomes exome capture data of 60 CLM, 66 MXL, and 55 PUR individuals because the high coverage reduces sequencing artifacts and synonymous sites are less affected by selection compared to non-synonymous sites.  A complete model with admixture would require at least one European, one African, and three Native American populations, which is beyond the 3-population limit of $\partial a\partial i.$ We therefore wish to focus on variants within Native American backgrounds.

Unfortunately, trio-phased sequencing data was not available for most samples. Because of phasing uncertainty, the actual ancestry assignment for variants at ancestry-heterozygous loci is uncertain.  To overcome this, we introduce a \emph{negative ascertainment} scheme, in which we only consider variable sites that have not been observed in any of the non-Native populations in the 1000 Genomes data set. The effect of this ascertainment scheme is to remove the majority of variants that predate the split of Native Americans from the rest of the populations. An additional benefit of this approach is that the impact of European and African tracts incorrectly assigned as Native American will be substantially reduced. We hypothesized that the effect of negative ascertainment could be approximately modeled by a strict bottleneck at the Native/non-Native split time. This was confirmed through simulations (see \ref{suppmet}).  

\begin{figure}
\includegraphics{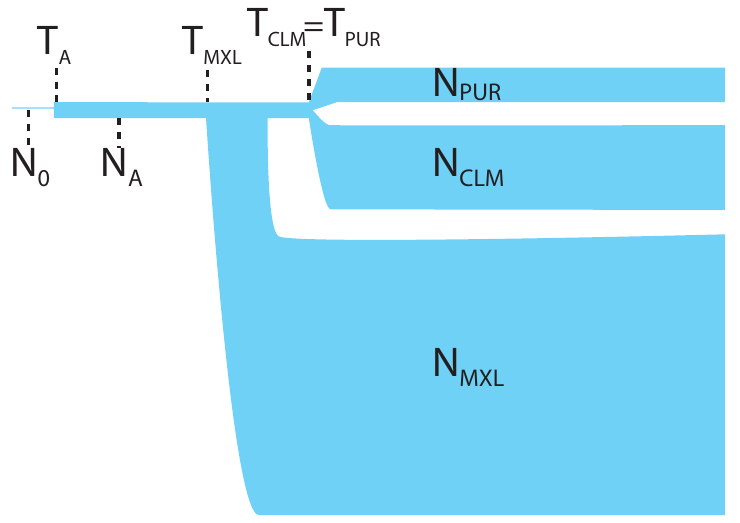}
\caption{\label{modelpic} An illustration of the maximum likelihood demographic model for the Native American ancestors to the CLM, MXL, and PUR panels. Parameter values are provided in Table \ref{paramest}. The ordering of the split shown (i.e., MXL splitting first) maximized the likelihood, but among the bootstrap replicates all three orders were observed.}
\end{figure}

We considered a simple 3-population demographic model starting with a constant population $N_0$. At time $T_A$ the population size changes to $N_A$. From this population of size $N_A$, population $i$ diverged with size $N_i$ at time $T_i$ and populations $j$ and $k$ diverge at a later time $T_j$ with respective sizes $N_j$ and $N_k$. {\color{black}We considered all three split orderings, with $i\in \{\mbox{CLM},\mbox{MXL},\mbox{PUR}\}$. In the optimal model, illustrated on Figure \ref{modelpic}, we have $i=\mbox{MXL}$, $j=\mbox{CLM}$, $k=\mbox{PUR}$. } 

\begin{table*}
\centering

\begin{tabular*}{\hsize}{@{\extracolsep{\fill}} ccc}
Parameter& Inferred value &$95\%$ CI\\ 
\hline
$N_A$&$514$&$316-2,264$\\ 
$N_{MXL}$&$62,127$&$48,824-127,897$ \\ 
$N_{CLM}$&$8,653$&$6,603-11,257$ \\ 
$N_{PUR}$&$1,922$&$1,456-2,748$\\ 
$T_A (y)$&$16,000$&\\ 
$T_{MXL} (y)$&$ 12,219 $&$11,157-12,595$\\
$T_{CLM} (y)$&$ 11,727 $&$9,807-12,822$\\
$T_{PUR} (y)$&$ 11,727 $&$9,807-12,742$\\
$\mu (\frac{10^{-8}}{ bp-gen})$&$ 1.44 $&$1.32-1.53$\\
\hline
\end{tabular*}
\caption{ Parameter estimates  for the model displayed on Figure \ref{modelpic}, assuming a bottleneck at the foundation of the Americas 16,000 years ago. 
  \label{paramest} }
\end{table*}

This model is a vast oversimplification of the historical demographic processes. However, given the limited statistical power to reconstruct time-dependent demographic histories using allele frequency data (e.g. \cite {Myers:2008fc}), such simple models with step-wise constant population sizes provide useful coarse-grained pictures of human demography. The population sizes in this model are effective population sizes: they are the size of  Wright-Fisher populations that best explain the observed patterns of polymorphism. They differ from census sizes because of population size fluctuations, overlapping generations, sex bias, offspring number dispersion, and other departures from the Wright-Fisher assumptions. The ratio $N_A/N_0$ is expected to converge to large values to reflect both the negative ascertainment scheme (see Methods) and the expansion post-founding of the Americas. The current data does not enable us to model these two effects separately, so the recovery time $T_A$ can be thought of as an interpolation between the two events. When performing likelihood optimization, $N_A/N_0$ tended to slowly increase without bound. Beyond a value of 100, this had minimal impact on the likelihood function and other parameter estimates. We therefore fixed this value to $N_A/N_0=100$ to facilitate optimization and prevent numerical instabilities.  All other parameters, and the order of population splits, were chosen to maximize the model likelihood.

We find dramatic differences in the inferred population sizes of the Native Ancestors to the MXL, CLM, and PUR (see Table \ref{paramest}), with the MXL showing by far the largest effective population size at 64,000, $7$ times larger than the CLM and 32 times larger than the PUR. Given the many sources of uncertainty and model limitations, these ratios are in good qualitative agreement with pre-Columbian populations estimated at 14M in central Mexico \cite{Salzano:2005wc}, 3M in Colombia\cite{Salzano:2005wc}, and somewhat over 110,000 in Puerto Rico \cite{Moscoso:2008ti}.  This could largely be a coincidence, given that the Native ancestors to the MXL and CLM were not panmictic populations over present-day political divisions. Another possible explanation for the differences in effective population sizes is a serial founder model after the crossing of Beringia: CLM and PUR would have experienced stricter and longer bottlenecks compared to MXL due to greater distances traveled from Beringia. The crossing to Puerto Rico is likely to have introduced intense bottlenecks in PUR, resulting in a smaller recent effective population size. 

The model suggests that PUR and CLM ancestral populations did not share serial founding events past the split with the MXL ancestors and split well before the expected arrival of the Arawak people of the Caribbean. Indeed, the first and second split times  ($T_i$ and $T_j$, respectively) are remarkably close to each other, with $T_i/T_j= 1.04$ (bootstrap $95\%$ CI: $1.01-1.18$, see \ref{suppmet}, Figure \ref{boothists}, and Table \ref{paramest}). This corresponds to a difference of about 500 years, 12,000 years ago. In fact, the splits are so close that it is impossible to distinguish which population split first, with bootstrap instances supporting all three orderings: the Ta\'ino ancestry does not appear much more closely related to either CLM or MXL Native ancestors. This is also consistent with the PCA results shown in Figure \ref{PCA}, showing a clear distinction between Native American groups in eastern and western Colombia.

Despite strong historical evidence for extensive population bottlenecks suffered by Native American populations following the arrival of Europeans  \cite{Dobyns:1966un}, we could not detect the presence of such bottlenecks through allele frequency analysis. However, the presence of such bottlenecks may affect our interpretation of effective population sizes. To quantify this, we fixed the timing and magnitudes of bottlenecks using non-genetic sources, and re-inferred model parameters. Dobyns  \cite{Dobyns:1966un} proposed a maximum population reduction of $95\%$ in the Native American population after European contact, but this number is expected to vary from location to location. Because we are studying admixed populations, the size of the bottleneck is related to the number of individuals that contributed to the admixed population, thus Dobyns' estimate may not apply.  In PUR, where the decline was particularly abrupt, we considered a decline of $98.5\%$ spanning $250$ years (see \ref{suppmet}). We found that inferred parameters were little affected by the existence of such a bottleneck, with the exception of the effective population size in the pre-bottleneck PUR population, which would be 3.9 times larger than in the no-bottleneck model. Assuming an additional bottleneck in the CLM population led to similar 4-fold increase in inferred pre-bottleneck CLM population size, with little effect on inferred split times. These are significant effects, but are less than the inferred differences in effective population sizes. Thus, in the absence of extreme differences in the recent bottlenecks experienced by the three populations, the observed differences in population sizes likely point to differences in pre-Columbian demography.

\begin{figure}
\scalebox{0.7}{\includegraphics{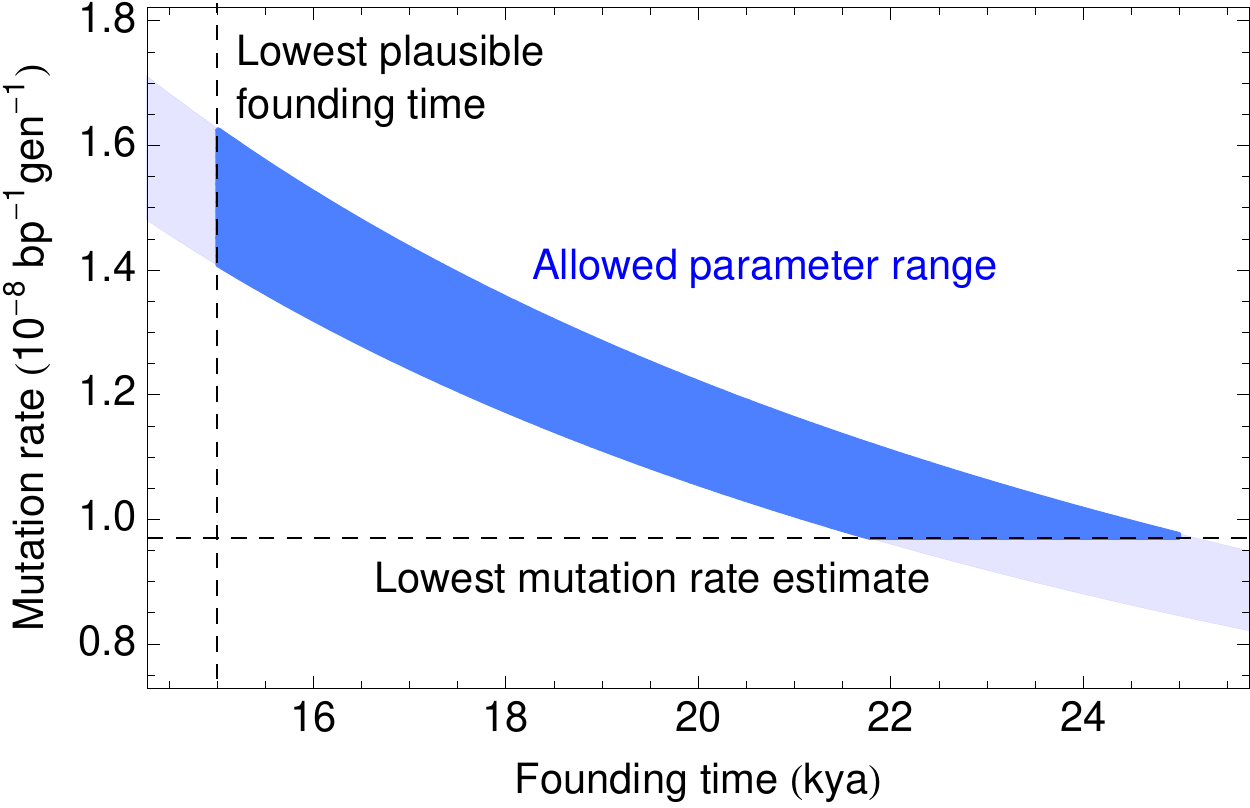}}
\caption{\label{mutbot} Plausible parameter range for the human mutation rate and the founding time of the Native American populations. The shaded blue area is the $95\%$ confidence interval from the current analysis. The horizontal line shows the lowest mutation rate estimate from \cite{Scally:2012fe}, and the vertical line shows the lowest plausible date for the founding of the ancestral Native American populations according to \cite{Dillehay2008}.  The plausible region, given by the overlap of the three areas, would correspond to a mutation rate of $0.97-1.6 \times 10^{-8} bp^{-1}gen^{-1}$ and a Native American founding time $15-24 kya$.  }
\end{figure}

By calibrating our results using  $T_A=16\mathrm{kya}$, towards the most recent end of the range of plausible values for the peopling of the Americas (see e.g., \cite{Dillehay2008} and references therein),  we find a mutation rate of $1.44\times 10^{-8}bp^{-1}gen^{-1}$ (bootstrap $95\%$ CI: $1.32-1.53\times 10^{-8}bp^{-1}gen^{-1}$), within the range of recently published human mutation rates \cite{Scally:2012fe}. The narrowest confidence interval reported in \cite{Scally:2012fe} was $1.05-1.5 \times 10^{-8}bp^{-1}gen^{-1}$, obtained from a de novo exome sequencing study \cite{Sanders:2012br}. Our sampling confidence interval is narrower than this value, but the main source of uncertainty here is the degree to which the bottleneck in our model reflects the bottleneck at the founding of the Americas, or the earlier split with the ancestors to the Chinese (CHB) and Japanese (JPT) sample, as well as uncertainty with respect to the timing of these two events  (see Figure \ref{mutbot}). 
The effect of changing the founding time or mutation rate assumptions would be to scale all parameters and confidence intervals according to  $T \propto N \propto 1/\mu.$ Thus the absolute uncertainty on individual parameters is larger than the sampling uncertainty suggests.

\subsection*{Estimating Native American allele frequencies}

There is scarce publicly available, genome-wide data about Native American genomic diversity. The 1000 Genomes dataset offers the opportunity to provide a diversity resource for Native American genomics by reconstructing the genetic makeup of Native American populations ancestral to the PUR, CLM, and MXL. This is particularly interesting in the case of the Puerto Rican population, where such reconstruction may be the only way to understand the genetic make-up of the pre-Columbian inhabitants of the Islands. Using the expectation maximization method presented in the Methods section, we estimated the allele frequencies in the Native-American-inferred part of the genomes of the sequenced individuals. These estimates are available at \url{http://genomes.uprm.edu/cgi-bin/gb2/gbrowse/}. 

\begin{figure}
\scalebox{0.6}{\includegraphics{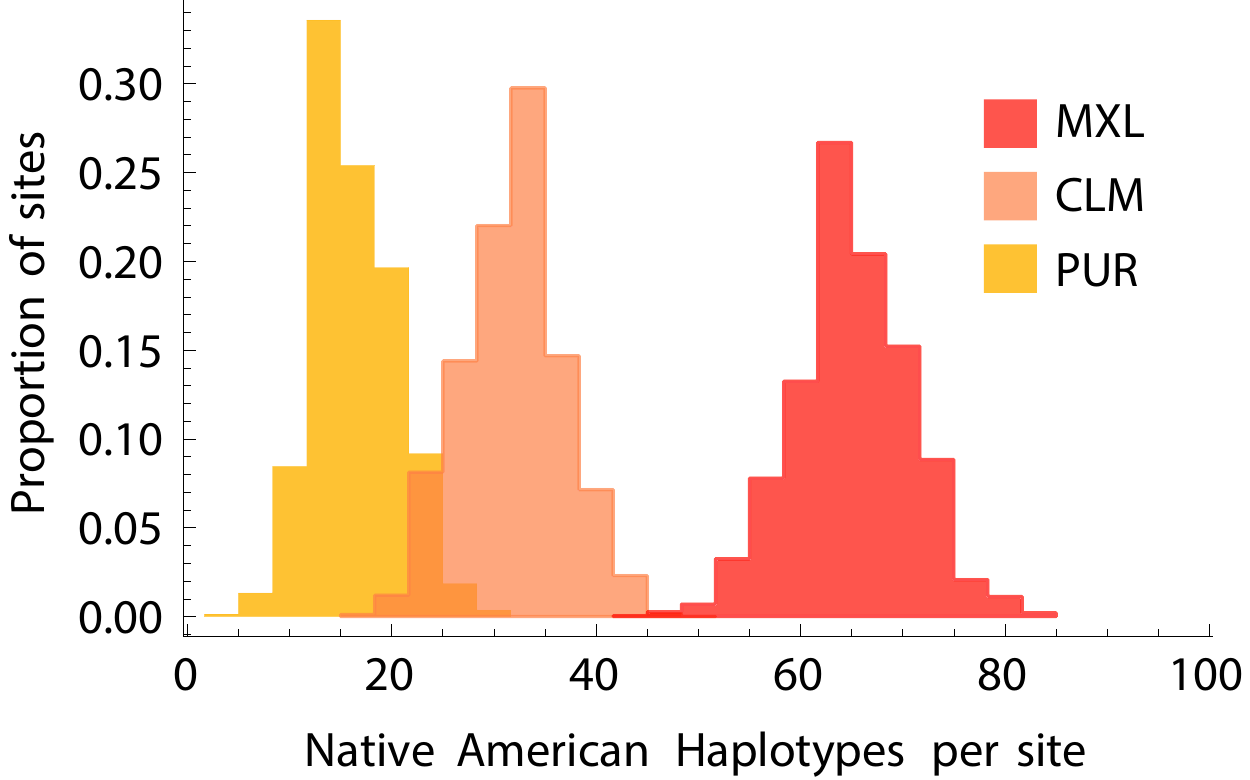}}
\scalebox{0.6}{\includegraphics{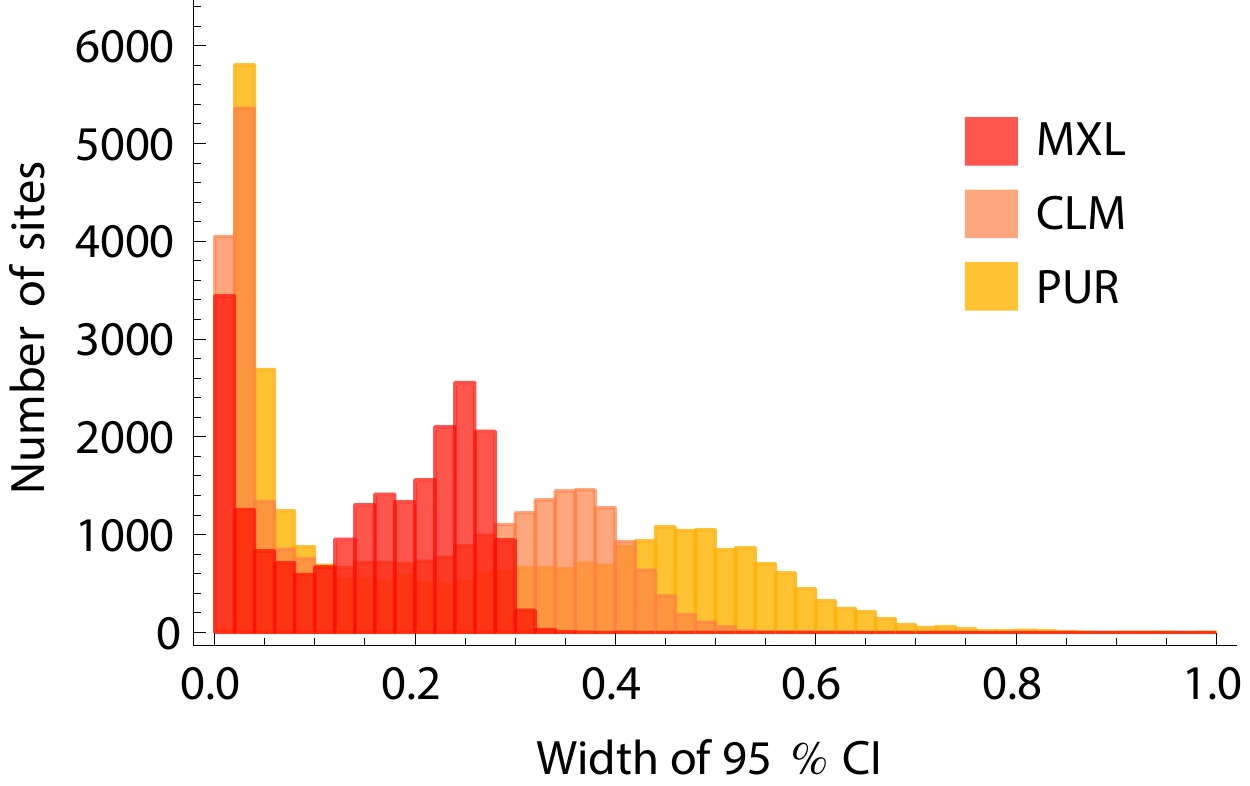}}
\caption{\label{confint} (a) Number of inferred Native American haplotypes per site, out of 120 CLM, 132 MXL, and 110 PUR haplotypes. (b) Distribution of confidence intervals widths for allele frequency estimations among the exomic Native American segments of the three panels.}
\end{figure}

Figure \ref{confint} shows the distribution of the number of Native American haplotypes per site and the resulting confidence intervals for allele frequency in each population for exome capture target regions. Absolute confidence intervals are narrow for rare variants, and reach a maximum for SNPs at intermediate frequency; the leftmost peak in the bimodal distribution corresponds to the large number of rare variants, whereas the right most peak encompasses a broader range of frequencies. 

Focusing on the $29,354$ variants with observations in all populations and within the exome capture regions, where coverage and accuracy were highest, the most significantly different among Native groups is rs11183610 on chromosome 12, with an estimated frequency of $0.49~~(95\%: 0.38-0.58)$ in MXL Native ancestry, $0.011~~(95\%: 0.00-0.12)$ in CLM Native ancestry, and $0.28~~(95\%: 0.02-0.49)$ in PUR Native Ancestry.  The MXL-PUR difference remains significant after Bonferroni correction (bootstrap $p=0.001$, see Methods). The bulk of the differentiation among populations is likely due to genetic drift, but such sub-continental ancestry informative markers are also interesting candidates for further selection scans.

\section*{Discussion}

{\color{black} The bottleneck at the founding of the Americas provides a unique opportunity to obtain precise estimates of the human autosomal mutation rate, as reported in Table \ref{paramest} and Figure \ref{mutbot}.  One remaining challenge in interpretation is whether the `founding time' studied here corresponds to the bottleneck at the founding of the Americas, or the split time of the Native Americans with the Asian populations. Fortunately, this uncertainty can be addressed by sequencing either trio-phased populations from the Americas, or individuals of Native American ancestry without large amounts of recent European and African ancestry. In either case, the dramatic events that led to the initial peopling of the Americas, together with the early dates of South American archaeological sites, provides us with estimates of the human mutation rate that are more precise than pedigree-based estimates. A more thorough study of the robustness of these estimates to model assumptions is therefore desirable. }

We find substantially larger effective population size in Mexico than in the other two populations through IBD-based and allele-frequency based estimates. These methods are sensitive to different time-scales: IBD analysis largely reflects post-Columbian events, as evidenced by the large number of mixed ancestry IBD segments in Figure \ref{IBDanc}(a). Allele frequencies reflect older events as well, and we showed that recent bottlenecks alone are unlikely to be responsible for the much larger effective MXL population size. To interpret the population size differences, we must consider the recent histories of the populations studied here. The MXL panel was recruited in Los Angeles among Mexican-American individuals, who may come from different regions in Mexico, a much wider geographical region than Puerto Rico, thus likely more populated. A natural question is whether the larger effective population sizes in MXL reflect a large panmictic population in Mexico, or a large number of small, previously isolated populations.  Figure \ref{PCA} and references \cite{Gorostiza:2012gp, Reich:2012cs} provide compelling evidence that there is substantial population structure within Native groups of Mexico. However, Figure \ref{PCA} also shows that the Native component of the MXL forms a relatively homogeneous cluster together with populations from southern Mexico. The much larger Native populations in central and southern Mexico are likely to have contributed the most to the Native American ancestry of Mexican mestizos, and thus Mexicans-Americans. Even though the MXL may have ancestors in different parts of Mexico, their Native genetic origins likely reflect the demographic history of the areas in Mexico with the highest Native American population sizes.   


Because Puerto Rico is an island, building a relatively complete population genetic model for the population may be more tractable. Clearly, our model of a single idealized pre-Columbian Native American, European, and African populations, joining to form a panmictic admixed population, is an oversimplification. African and European ancestry proportions vary along the island \cite{via:2011hs} and eastern parts of Puerto Rico, with elevated proportions of African ancestry, are underrepresented in this study.  By contrast, we do not have evidence for variation in the amount or composition of the Native American ancestry across the island, and it is likely that the conclusions about the pre-Columbian Native American fraction of the population are robust to sampling ascertainment. Interestingly, we find that the distribution of ancestry tract length in a sample of individuals of Puerto Rican descent in south Florida gave very similar results, despite different location, sequencing platform, and local ancestry inference method \cite{MorenoEstrada:2013wx}. Historical gene flow inference using individuals of Colombian descent in south Florida provided comparable estimates of the time of admixture onset, but different patterns of recent gene flow--as is typical in demographic inference, inference of recent events is more sensitive to population structure. 

Our analyses largely rely on accurate estimates of local ancestry patterns along the genome obtained through \textsc{RFMix}. This method has been shown to provide more than $95\%$ accuracy on three-way admixture using comparable reference panels \cite{Maples:2013fia}, an accuracy level that enables accurate estimation of genome-wide diversity \cite{Kidd:2012fi}. To ensure that our results are robust to residual errors, we further took into account the difficulty of calling short ancestry tracts in our migration estimates, and performed negative ascertainment of non-Native American alleles in the demographic inference. Some of these results can be independently verified by independent sequencing of contemporary or ancient individuals with more uniform ancestry. However, understanding the genetic history of admixed populations will continue to rely on statistically picking apart the contributions of different ancestral populations, and the development of improved statistical methods, particularly for admixture that is ancient or between closely related populations, remains highly desirable.

The genetic heterogeneity in continental ancestry proportions among populations of the Americas is well appreciated \cite{Bryc:2010bp, Wang:2008cg, Bedoya:2006ei}. Our results emphasize more fine-scale aspects of this diversity: because of the similarity between European founders of different populations and the high divergence among the Native American ancestors, populations that appear similar under classical tests such as $F_{ST}$ or principal component analysis may still harbor population specific Native American haplotypes that must be carefully accounted for when performing rare-variant association testing in cosmopolitan cohorts. Similarly, the choice of a replication cohort for an identified risk variant should be guided by the ancestral background on which the variant is  found. The PUR may be an excellent replication cohort for a result found in CLM if the background is European. If the background is Native American, a different cohort with related Native Ancestry would likely be much more appropriate. Understanding the genetics of the different ancestral populations of the Americas, and the relatedness among these ancestral groups, will therefore facilitate the development of association methods that account for and take advantage of this rich diversity.   
\section*{Methods}

\subsection*{Negative ascertainment}
\label{theory}
Ideally, we would have been able to directly model the joint site-frequency spectrum (SFS) of all the ancestral populations to the PUR, CLM, and MXL. However, because we are interested in distinguishing the Native American ancestries to the three populations, this would require modeling at least 5 populations, which is beyond the scope of current methods. We would like to use the inferred local ancestry to focus on the Native American ancestry only, but this is difficult because most Native American haplotypes are in segments heterozygous for ancestry. Because of phasing errors, allele-specific ancestry can be incorrectly assigned. To minimize the impact of such mis-assigned ancestry and to ensure that we focused on variants of genuine Native American ancestry, we discarded all variants observed in 1000 Genomes individuals of African, European, and Asian ancestry, as well as variants observed in Hispanic/Latino populations in segments with no Native American ancestry inferred. 

We then considered all remaining variable sites that were assigned Nat/Nat diploid ancestry and Nat/Eur ancestry, and calculated the expected frequency distribution under the assumption of perfect negative ascertainment, that is, that all remaining variants were on the Native American background. Because the European backgrounds are expected to carry a number of singletons, this would result in an overestimate of the number of singletons in the Native Ancestry. Fortunately, this bias is easy to estimate empirically: we first choose $s_E$ segments of Eur/Eur ancestry to mimic the $2s_E$ European haplotypes in our sample. After performing the negative ascertainment scheme on these genotypes, we can directly estimate the bias in the negative ascertainment scheme. In practice, this correction is very low except for singletons, as expected. The number of excess singletons was 129 for CLM, 73 for PUR, and  40 for MXL. The largest non-singleton correction is 1.3 for doubletons in CLM.

Because negative ascertainment removes a significant proportion of the variants that were present at the Native American split from other populations, we hypothesized that this effect could be well-approximated by a severe bottleneck at the time of split between non-Native and Native American ancestry.

\begin{figure}
\scalebox{0.6}{\includegraphics{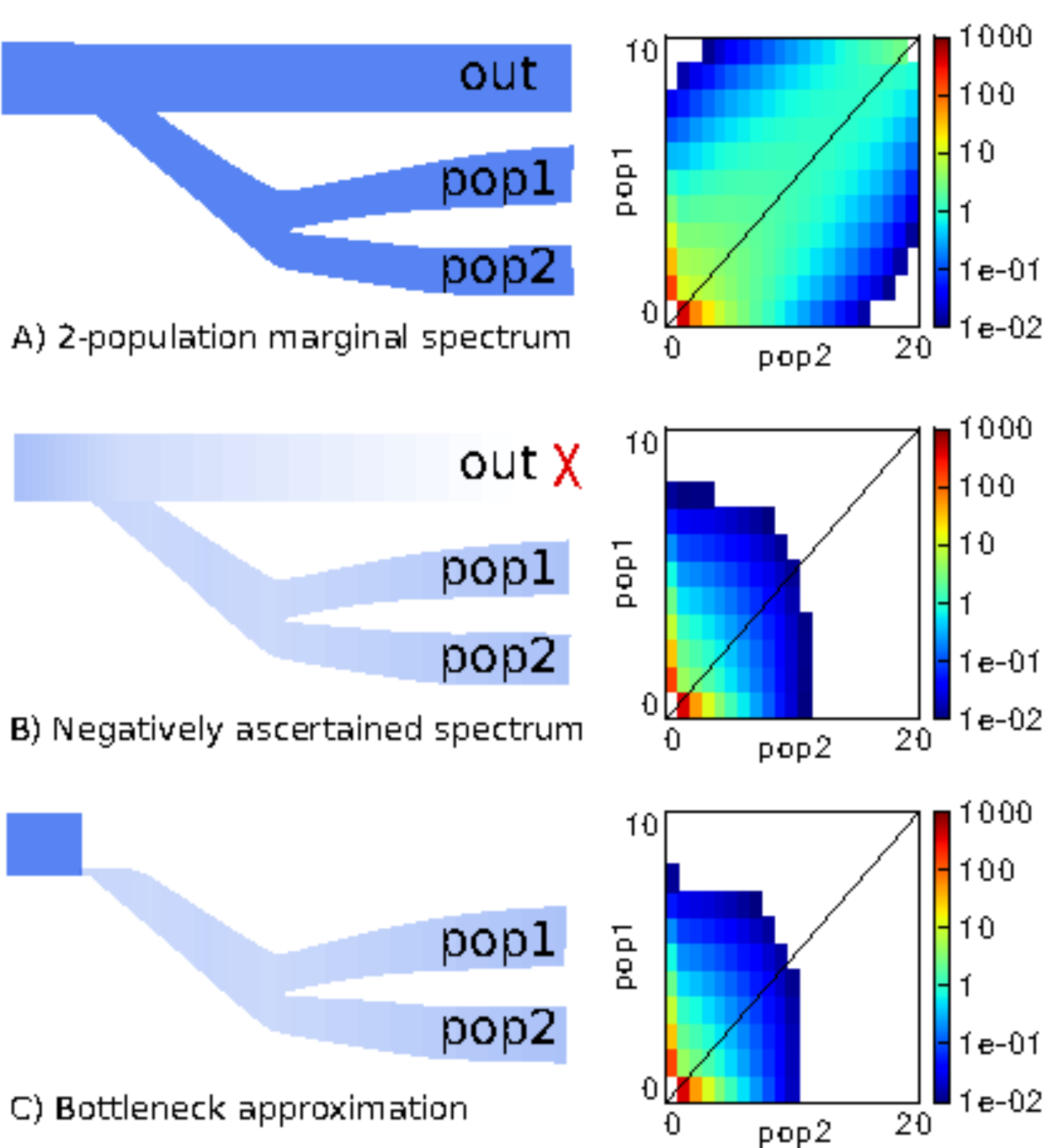}}
\caption{\label{simulpinch} Illustration of the negative ascertainment scheme, with simulation. (a) A basic three population model, showing the joint site-frequency spectrum for populations 1 and 2 as a heat map. (b)  Conditioning on variants not being observed in the out-population results in a SFS skewed towards rare variants. (c) A quantitatively similar effect can be obtained by introducing a drastic bottleneck at the root of the tree and considering only two populations.}
\end{figure}

Figure \ref{simulpinch} provides a simulated example, wherein a marginal spectrum (top) is compared to a spectrum negatively ascertained using 100 diploid individuals from the `outgroup' population (middle) and to a  bottleneck approximation equivalent (bottom). More quantitatively, we simulated a two-populations sample diverged  12.1kya, and negatively ascertained using a population diverged at 16.5 kya, and attempted to model this as a two-population model with an early bottleneck. The inferred bottleneck timing was within $3\%$ of the split time with the outgroup, and the three population sizes and split time between populations 1 and 2 were within $1.2\%$ of the correct value.   These biases are well within the acceptable range given other biases and uncertainties. 


  
 \subsection*{Allele frequencies in Native American segments}

We wish to estimate the allele frequencies at each site among segments of Native American origin, but we have to contend with a finite sample and inaccurate phasing. We therefore choose to model the underlying population frequency $\mathbf{f}$ across all populations using Bayes rule

 \begin{equation}
 P(\mathbf{f}|D,R)=\frac{P(D |\mathbf{f},R) P(\mathbf{f}| R)}{\int d\mathbf{f'} P(D |\mathbf{f'},R) P(\mathbf{f'}|R)},
 \label{Estep} \end{equation}
  where $D$ is the observed genotype data, $D\in \{00,01,11\}$, and $R$ is the diploid local ancestry calls (e.g., $R\in \{AA,AB,BB\}$ for populations A and B). From this distribution we can calculate expected frequency and confidence intervals.
  We report inferred frequencies and confidence intervals at non-monomorphic sites.  
   
 To estimate $P(D |\mathbf{f},R)$, we write $\mathbf{f}=\{f_A,f_B\}$ as the frequencies of the non reference allele in populations $A$ and $B$. We have  $P(01|\mathbf{f}, AB)=f_A(1-f_B) + f_B (1-f_A)$,  for ancestry and genotype heterozygous segments,  $P(11|\mathbf{f}, AB)=f_A f_B$, and so forth. 
 To estimate $P(\mathbf{f}| R)$, we first observe that because we are considering population frequencies, rather than sample frequencies, $\mathbf{f}$ is independent of $R$: $P(\mathbf{f}| R)\simeq P(\mathbf{f}) $. This suggests the use of a self-consistent, expectation-maximization procedure. We estimate the underlying frequency distribution as

\begin{equation}
P(\mathbf{f})=\frac{\sum_{s} P(\mathbf{f_s}|D_s, R_s)}{\#s},
\label{Mstep}
\end{equation}
the sum over the estimated probabilities at each site. We can thus iterate Equations \eqref{Estep} and \eqref{Mstep} until self-consistency is reached to estimate both allele frequency distributions and single-site allele frequencies in each population. 

A final caveat is that the sum runs over all sites, including monomorphic ones. If we only observe the subset of sites that are polymorphic, an additional step is needed. If $\#n$ is the number of monomorphic (unobserved) sites (denoted as $M$), and  $\sum'$ represents the sum over polymorphic sites,  we have 
  \begin{equation}
  \begin{split}
  P(\mathbf{f})&\simeq \frac{\sum_{s}' P(\mathbf{f}|D)+\#n P(\mathbf{f}|M)} {\#s}\\
  &\simeq \frac{\sum'_{s} P(\mathbf{f}|D)}{\#s}+P(M) P(\mathbf{f}|M)\\
 &= \frac{\sum'_{s} P(\mathbf{f}|D)}{\#s} +P(M| \mathbf{f}) P(\mathbf{f})
 \end{split}
 \end{equation}
 and, therefore, 
 $$P(\mathbf{f})=\frac{\sum'_{s} P(\mathbf{f}|D)}{\#s \left(1-P(M| \mathbf{f})\right)}.$$ 
 
 Intuitively, we are correcting for the proportions of sites at every frequency that might have gone undetected. Results are reported using 20 EM iterations, for sites where all individuals had both ancestry and genotype calls, and data can be downloaded at \url{http://genomes.uprm.edu/cgi-bin/gb2/gbrowse/}.

 To test this method, we considered 84 diploid individuals,  each formed by drawing two chromosomes (without replacement) from 84 CEU and 84 YRI individuals, resulting in a simulated 50-50 admixture proportion. We considered 100,000 sites on chromosome 22, and performed the EM inference as described.   
 
 Among the 85677 sites that were found to be polymorphic, only 13 had a sample allele frequency departing from the $95\%$ confidence interval for the European ancestry, and 51 among the African ancestry. Confidence intervals encompass much more than $95\%$ of \emph{sample} allele frequencies, emphasizing that the width of the confidence interval largely reflects the uncertainty about the \emph{population} frequency given a fixed sample frequency, rather than the phasing uncertainty.

 \subsection*{Optimizing the demographic model}
Because the demographic model considered here does not involve migrations between Native groups, we considered the composite likelihood of three pairwise two-population allele frequency distributions, rather than the full three-population spectrum. This allows for much faster inference and better convergence of the numerical optimization. In principle, it also enables the joint inference of more than three populations. We showed through simulations that the use of a composite likelihood had an effect on inferred parameters that was much smaller than other sources of uncertainty. We used grids of 20,40, and 60 grid points per population, and projected Native American allele frequencies to sample sizes of 10 in PUR, 20 in CLM, and 40 in MXL.

\section*{Acknowledgements} 
\subsection*{Members of the 1000 Genomes project}
Altshuler DM, Durbin RM, Abecasis GR, Bentley DR, Chakravarti A, Clark AG, Donnelly P, Eichler EE, Flicek P, Gabriel SB, Gibbs RA, Green ED, Hurles ME, Knoppers BM, Korbel JO, Lander ES, Lee C, Lehrach H, Mardis ER, Marth GT, McVean GA, Nickerson DA, Schmidt JP, Sherry ST, Wang J, Wilson RK, Gibbs RA, Dinh H, Kovar C, Lee S, Lewis L, Muzny D, Reid J, Wang M, Wang J, Fang X, Guo X, Jian M, Jiang H, Jin X, Li G, Li J, Li Y, Li Z, Liu X, Lu Y, Ma X, Su Z, Tai S, Tang M, Wang B, Wang G, Wu H, Wu R, Yin Y, Zhang W, Zhao J, Zhao M, Zheng X, Zhou Y, Lander ES, Altshuler DM, Gabriel SB, Gupta N, Flicek P, Clarke L, Leinonen R, Smith RE, Zheng-Bradley X, Bentley DR, Grocock R, Humphray S, James T, Kingsbury Z, Lehrach H, Sudbrak R, Albrecht MW, Amstislavskiy VS, Borodina TA, Lienhard M, Mertes F, Sultan M, Timmermann B, Yaspo ML, Sherry ST, McVean GA, Mardis ER, Wilson RK, Fulton L, Fulton R, Weinstock GM, Durbin RM, Balasubramaniam S, Burton J, Danecek P, Keane TM, Kolb-Kokocinski A, McCarthy S, Stalker J, Quail M, Schmidt JP, Davies CJ, Gollub J, Webster T, Wong B, Zhan Y, Auton A, Gibbs RA, Yu F, Bainbridge M, Challis D, Evani US, Lu J, Muzny D, Nagaswamy U, Reid J, Sabo A, Wang Y, Yu J, Wang J, Coin LJ, Fang L, Guo X, Jin X, Li G, Li Q, Li Y, Li Z, Lin H, Liu B, Luo R, Qin N, Shao H, Wang B, Xie Y, Ye C, Yu C, Zhang F, Zheng H, Zhu H, Marth GT, Garrison EP, Kural D, Lee WP, Leong WF, Ward AN, Wu J, Zhang M, Lee C, Griffin L, Hsieh CH, Mills RE, Shi X, von Grotthuss M, Zhang C, Daly MJ, DePristo MA, Altshuler DM, Banks E, Bhatia G, Carneiro MO, del Angel G, Gabriel SB, Genovese G, Gupta N, Handsaker RE, Hartl C, Lander ES, McCarroll SA, Nemesh JC, Poplin RE, Schaffner SF, Shakir K, Yoon SC, Lihm J, Makarov V, Jin H, Kim W, Kim KC, Korbel JO, Rausch T, Flicek P, Beal K, Clarke L, Cunningham F, Herrero J, McLaren WM, Ritchie GR, Smith RE, Zheng-Bradley X, Clark AG, Gottipati S, Keinan A, Rodriguez-Flores JL, Sabeti PC, Grossman SR, Tabrizi S, Tariyal R, Cooper DN, Ball EV, Stenson PD, Bentley DR, Barnes B, Bauer M, Cheetham R, Cox T, Eberle M, Humphray S, Kahn S, Murray L, Peden J, Shaw R, Ye K, Batzer MA, Konkel MK, Walker JA, MacArthur DG, Lek M, Sudbrak R, Amstislavskiy VS, Herwig R, Shriver MD, Bustamante CD, Byrnes JK, De La Vega FM, Gravel S, Kenny EE, Kidd JM, Lacroute P, Maples BK, Moreno-Estrada A, Zakharia F, Halperin E, Baran Y, Craig DW, Christoforides A, Homer N, Izatt T, Kurdoglu AA, Sinari SA, Squire K, Sherry ST, Xiao C, Sebat J, Bafna V, Ye K, Burchard EG, Hernandez RD, Gignoux CR, Haussler D, Katzman SJ, Kent WJ, Howie B, Ruiz-Linares A, Dermitzakis ET, Lappalainen T, Devine SE, Liu X, Maroo A, Tallon LJ, Rosenfeld JA, Michelson LP, Abecasis GR, Kang HM, Anderson P, Angius A, Bigham A, Blackwell T, Busonero F, Cucca F, Fuchsberger C, Jones C, Jun G, Li Y, Lyons R, Maschio A, Porcu E, Reinier F, Sanna S, Schlessinger D, Sidore C, Tan A, Trost MK, Awadalla P, Hodgkinson A, Lunter G, McVean GA, Marchini JL, Myers S, Churchhouse C, Delaneau O, Gupta-Hinch A, Iqbal Z, Mathieson I, Rimmer A, Xifara DK, Oleksyk TK, Fu Y, Liu X, Xiong M, Jorde L, Witherspoon D, Xing J, Eichler EE, Browning BL, Alkan C, Hajirasouliha I, Hormozdiari F, Ko A, Sudmant PH, Mardis ER, Chen K, Chinwalla A, Ding L, Dooling D, Koboldt DC, McLellan MD, Wallis JW, Wendl MC, Zhang Q, Durbin RM, Hurles ME, Tyler-Smith C, Albers CA, Ayub Q, Balasubramaniam S, Chen Y, Coffey AJ, Colonna V, Danecek P, Huang N, Jostins L, Keane TM, Li H, McCarthy S, Scally A, Stalker J, Walter K, Xue Y, Zhang Y, Gerstein MB, Abyzov A, Balasubramanian S, Chen J, Clarke D, Fu Y, Habegger L, Harmanci AO, Jin M, Khurana E, Mu XJ, Sisu C, Li Y, Luo R, Zhu H, Lee C, Griffin L, Hsieh CH, Mills RE, Shi X, von Grotthuss M, Zhang C, Marth GT, Garrison EP, Kural D, Lee WP, Ward AN, Wu J, Zhang M, McCarroll SA, Altshuler DM, Banks E, del Angel G, Genovese G, Handsaker RE, Hartl C, Nemesh JC, Shakir K, Yoon SC, Lihm J, Makarov V, Degenhardt J, Flicek P, Clarke L, Smith RE, Zheng-Bradley X, Korbel JO, Rausch T, StŸtz AM, Bentley DR, Barnes B, Cheetham R, Eberle M, Humphray S, Kahn S, Murray L, Shaw R, Ye K, Batzer MA, Konkel MK, Walker JA, Lacroute P, Craig DW, Homer N, Church D, Xiao C, Sebat J, Bafna V, Michaelson JJ, Ye K, Devine SE, Liu X, Maroo A, Tallon LJ, Lunter G, Iqbal Z, Witherspoon D, Xing J, Eichler EE, Alkan C, Hajirasouliha I, Hormozdiari F, Ko A, Sudmant PH, Chen K, Chinwalla A, Ding L, McLellan MD, Wallis JW, Hurles ME, Blackburne B, Li H, Lindsay SJ, Ning Z, Scally A, Walter K, Zhang Y, Gerstein MB, Abyzov A, Chen J, Clarke D, Khurana E, Mu XJ, Sisu C, Gibbs RA, Yu F, Bainbridge M, Challis D, Evani US, Kovar C, Lewis L, Lu J, Muzny D, Nagaswamy U, Reid J, Sabo A, Yu J, Guo X, Li Y, Wu R, Marth GT, Garrison EP, Leong WF, Ward AN, del Angel G, DePristo MA, Gabriel SB, Gupta N, Hartl C, Poplin RE, Clark AG, Rodriguez-Flores JL, Flicek P, Clarke L, Smith RE, Zheng-Bradley X, MacArthur DG, Bustamante CD, Gravel S, Craig DW, Christoforides A, Homer N, Izatt T, Sherry ST, Xiao C, Dermitzakis ET, Abecasis GR, Kang HM, McVean GA, Mardis ER, Dooling D, Fulton L, Fulton R, Koboldt DC, Durbin RM, Balasubramaniam S, Keane TM, McCarthy S, Stalker J, Gerstein MB, Balasubramanian S, Habegger L, Garrison EP, Gibbs RA, Bainbridge M, Muzny D, Yu F, Yu J, del Angel G, Handsaker RE, Makarov V, Rodriguez-Flores JL, Jin H, Kim W, Kim KC, Flicek P, Beal K, Clarke L, Cunningham F, Herrero J, McLaren WM, Ritchie GR, Zheng-Bradley X, Tabrizi S, MacArthur DG, Lek M, Bustamante CD, De La Vega FM, Craig DW, Kurdoglu AA, Lappalainen T, Rosenfeld JA, Michelson LP, Awadalla P, Hodgkinson A, McVean GA, Chen K, Tyler-Smith C, Chen Y, Colonna V, Frankish A, Harrow J, Xue Y, Gerstein MB, Abyzov A, Balasubramanian S, Chen J, Clarke D, Fu Y, Harmanci AO, Jin M, Khurana E, Mu XJ, Sisu C, Gibbs RA, Fowler G, Hale W, Kalra D, Kovar C, Muzny D, Reid J, Wang J, Guo X, Li G, Li Y, Zheng X, Altshuler DM, Flicek P, Clarke L, Barker J, Kelman G, Kulesha E, Leinonen R, McLaren WM, Radhakrishnan R, Roa A, Smirnov D, Smith RE, Streeter I, Toneva I, Vaughan B, Zheng-Bradley X, Bentley DR, Cox T, Humphray S, Kahn S, Sudbrak R, Albrecht MW, Lienhard M, Craig DW, Izatt T, Kurdoglu AA, Sherry ST, Ananiev V, Belaia Z, Beloslyudtsev D, Bouk N, Chen C, Church D, Cohen R, Cook C, Garner J, Hefferon T, Kimelman M, Liu C, Lopez J, Meric P, O'Sullivan C, Ostapchuk Y, Phan L, Ponomarov S, Schneider V, Shekhtman E, Sirotkin K, Slotta D, Xiao C, Zhang H, Haussler D, Abecasis GR, McVean GA, Alkan C, Ko A, Dooling D, Durbin RM, Balasubramaniam S, Keane TM, McCarthy S, Stalker J, Chakravarti A, Knoppers BM, Abecasis GR, Barnes KC, Beiswanger C, Burchard EG, Bustamante CD, Cai H, Cao H, Durbin RM, Gharani N, Gibbs RA, Gignoux CR, Gravel S, Henn B, Jones D, Jorde L, Kaye JS, Keinan A, Kent A, Kerasidou A, Li Y, Mathias R, McVean GA, Moreno-Estrada A, Ossorio PN, Parker M, Reich D, Rotimi CN, Royal CD, Sandoval K, Su Y, Sudbrak R, Tian Z, Timmermann B, Tishkoff S, Toji LH, Tyler-Smith C, Via M, Wang Y, Yang H, Yang L, Zhu J, Bodmer W, Bedoya G, Ruiz-Linares A, Ming CZ, Yang G, You CJ, Peltonen L, Garcia-Montero A, Orfao A, Dutil J, Martinez-Cruzado JC, Oleksyk TK, Brooks LD, Felsenfeld AL, McEwen JE, Clemm NC, Duncanson A, Dunn M, Green ED, Guyer MS, Peterson JL.
 


 \bibliography{morebiblio,biblio}

\pagebreak

\clearpage
\pagestyle{empty}

\appendix
\setcounter{figure}{0}
\makeatletter
\renewcommand{\thefigure}{S\@arabic\c@figure}


\begin{figure}
\scalebox{0.25}{\includegraphics[]{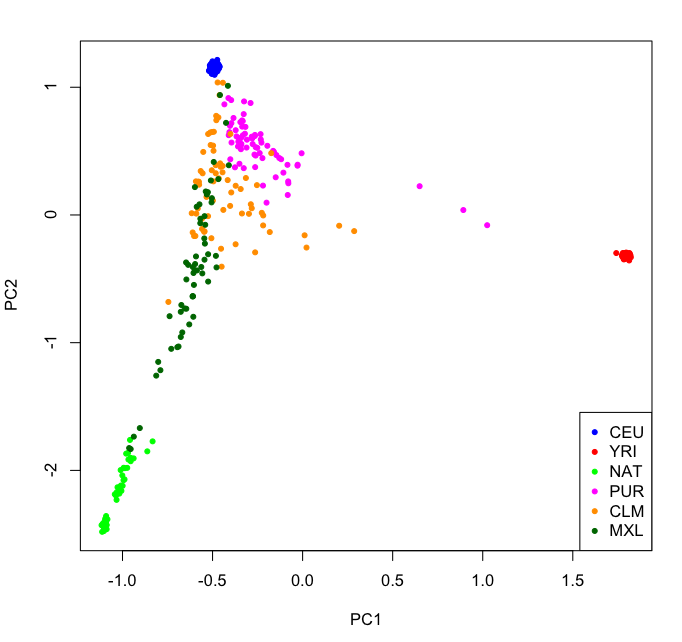}}
\caption{\label{regPCA} The first two principal components for 1000 Genomes populations, showing the distribution of admixed populations}
\end{figure}

\begin{figure}
\scalebox{0.4}{\includegraphics{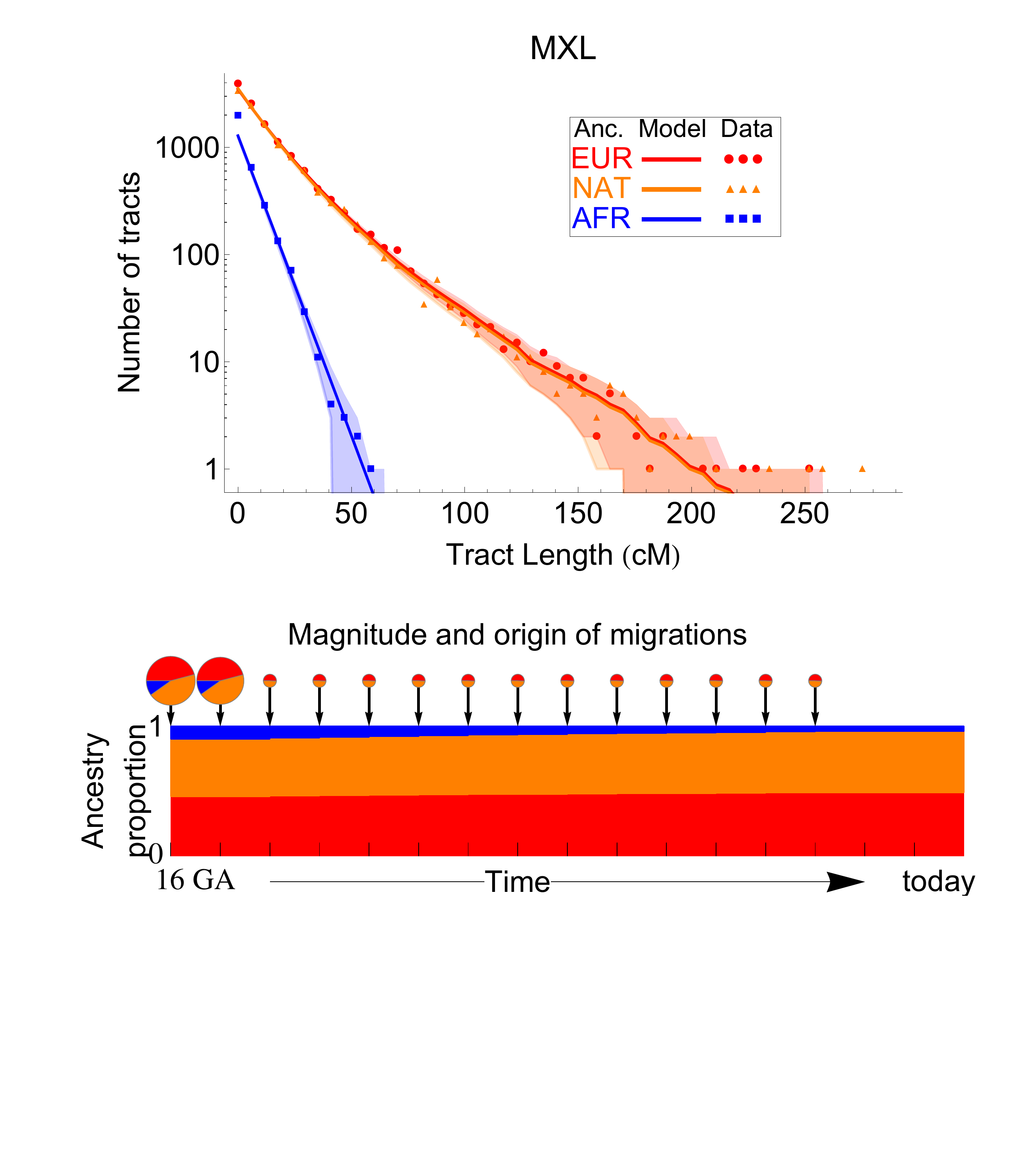}}
\caption{\label{MXLtracts} Ancestry tract length distribution in MXL compared to the predictions of the best-fitting migration model (displayed below). Solid lines represent model predictions and shaded areas are one-sigma confidence regions surrounding the predictions, assuming a Poisson distribution   \cite{Kidd:2012fi} }
\end{figure}

\begin{figure}
\scalebox{0.45}{\includegraphics{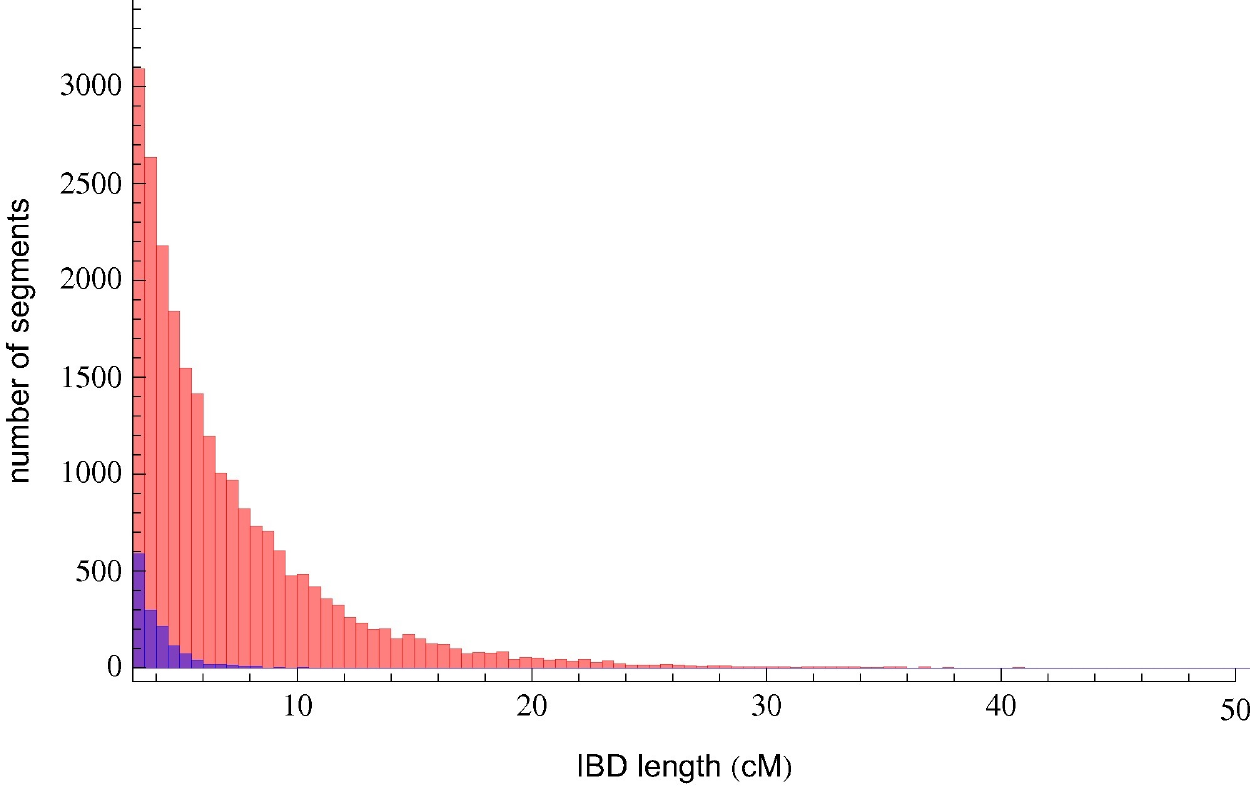}}
\caption{\label{histIBD}Distribution of IBD lengths within populations (red) and across populations (purple).}
\end{figure}

\begin{figure}
\scalebox{0.6}{\includegraphics{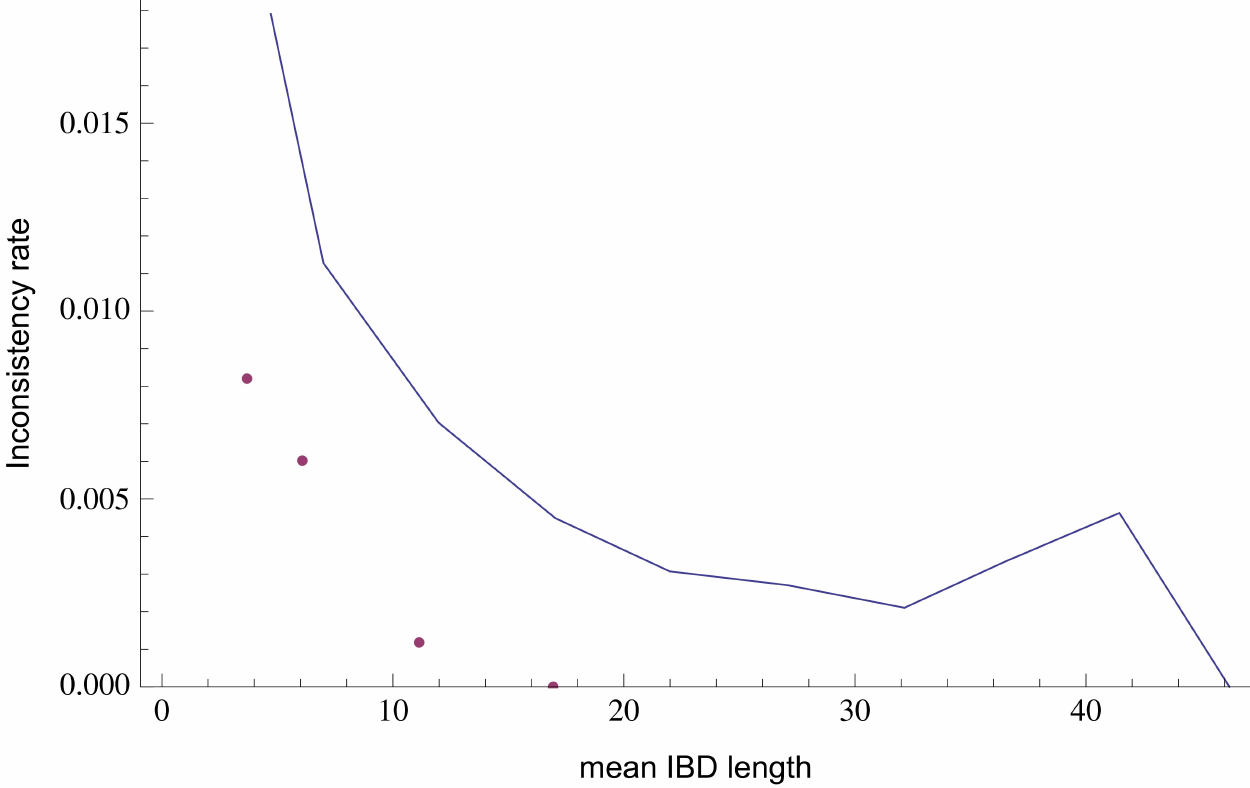}}
\caption{\label{IILANC} IBD inconsistency rate as a function of IBD length. Long IBD segments exhibit significantly fewer ancestry inconsistencies. The line represents within-population IBD, the red dots represents across-population IBD. }
\end{figure}

\begin{figure}
\scalebox{0.4}{\includegraphics{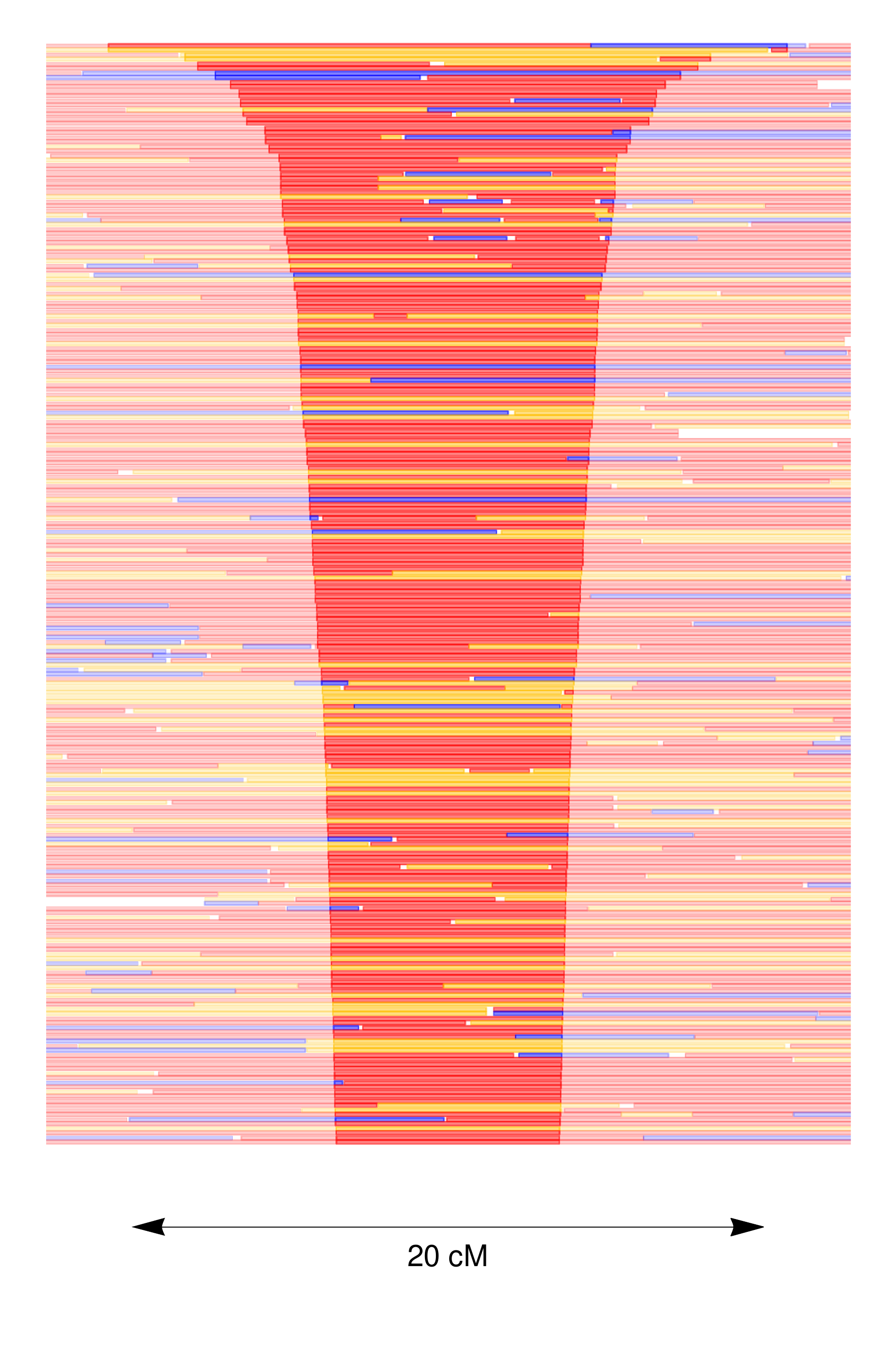}}
\caption{\label{altcontrol} Ancestry assignments in a control formed by taking the non-IBD matching haplotypes at loci where the alternate haplotype are IBD }
\end{figure}


\begin{figure}
\scalebox{0.5}{\includegraphics{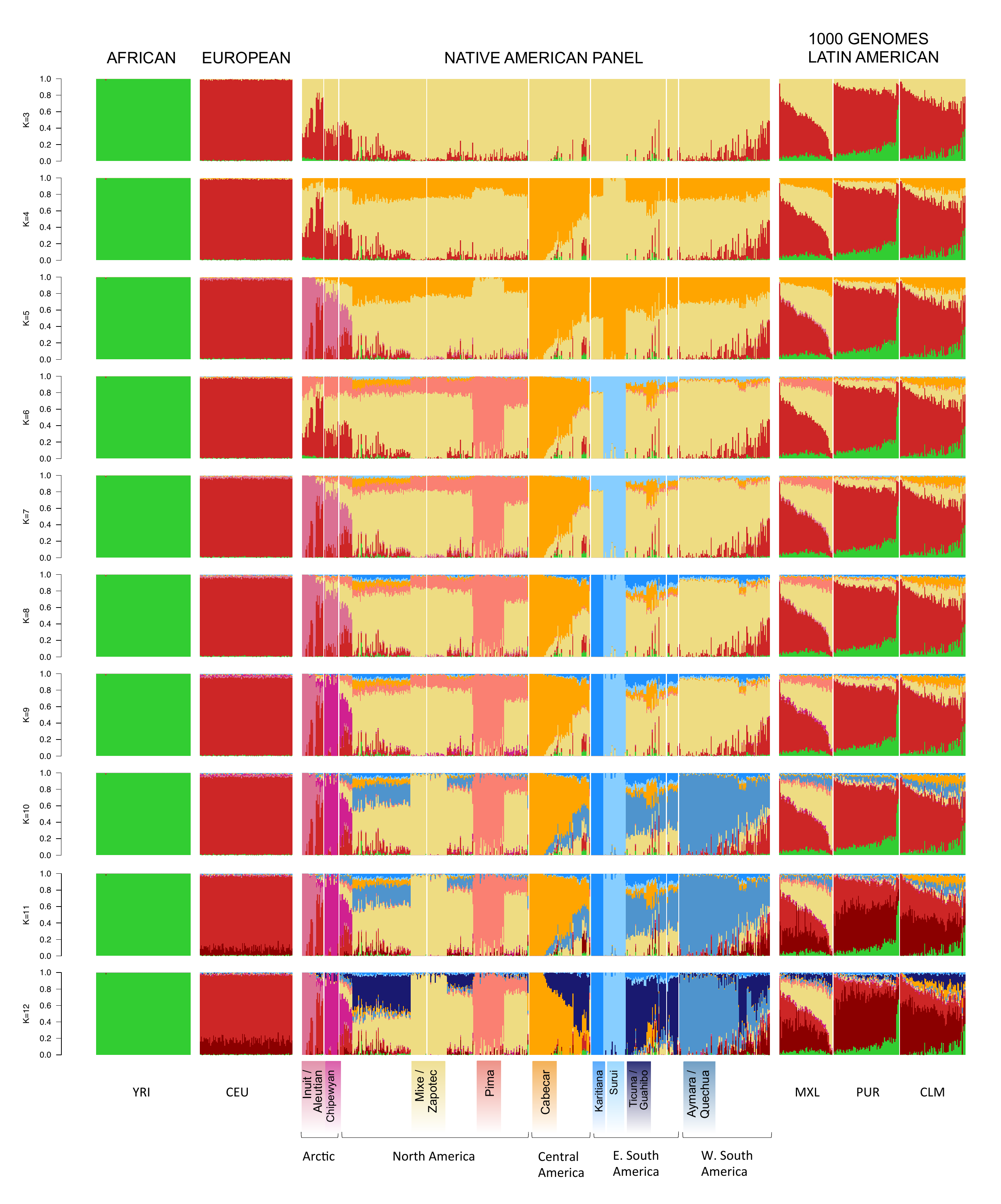}}
\caption{\label{Admixture312} Results of {\sc admixture} analysis with K=3 to K=12, with Native American populations grouped by geographic origin.  }
\end{figure}

\begin{figure}
\scalebox{0.4}{\includegraphics{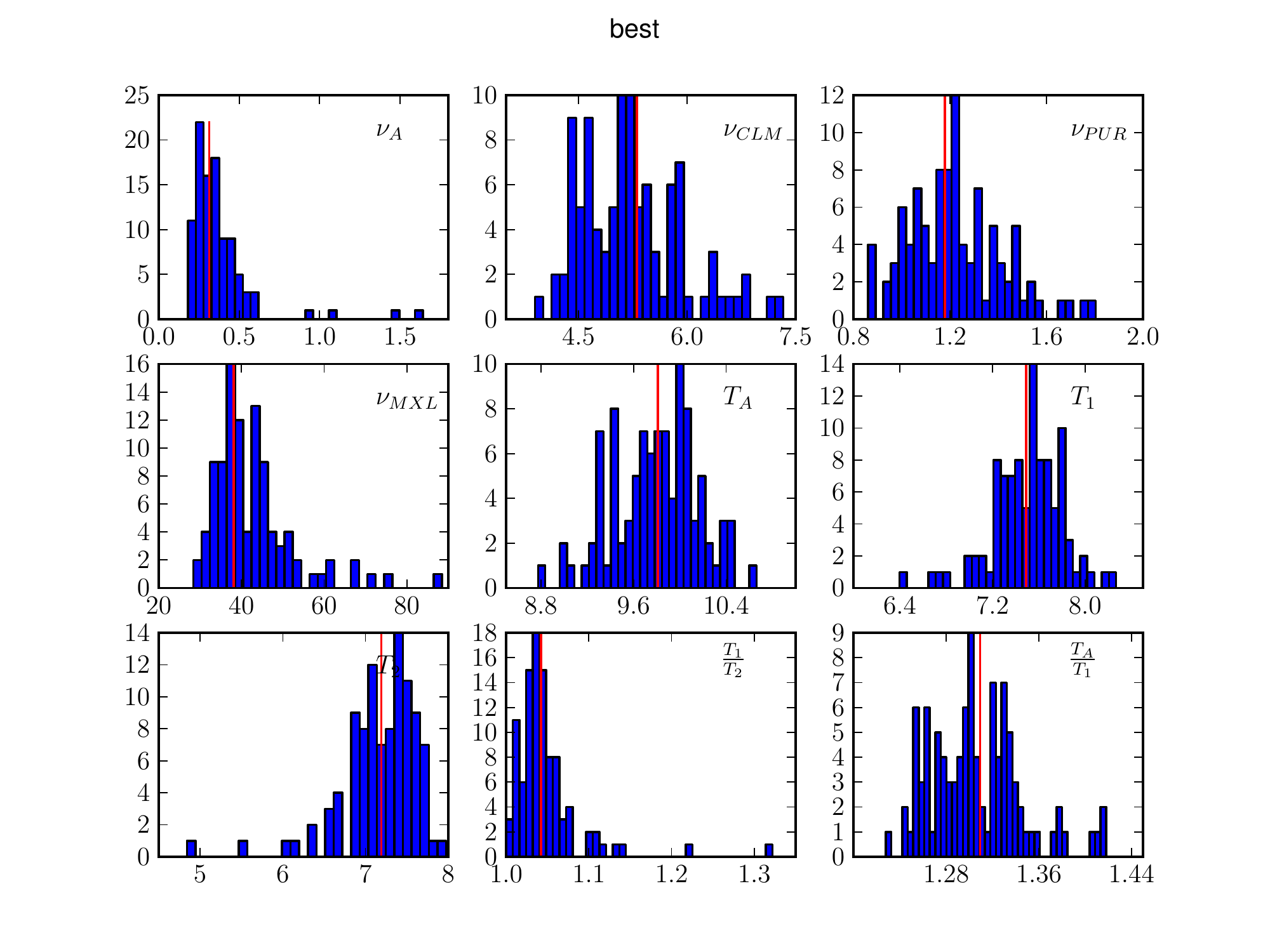}}
\scalebox{0.4}{\includegraphics{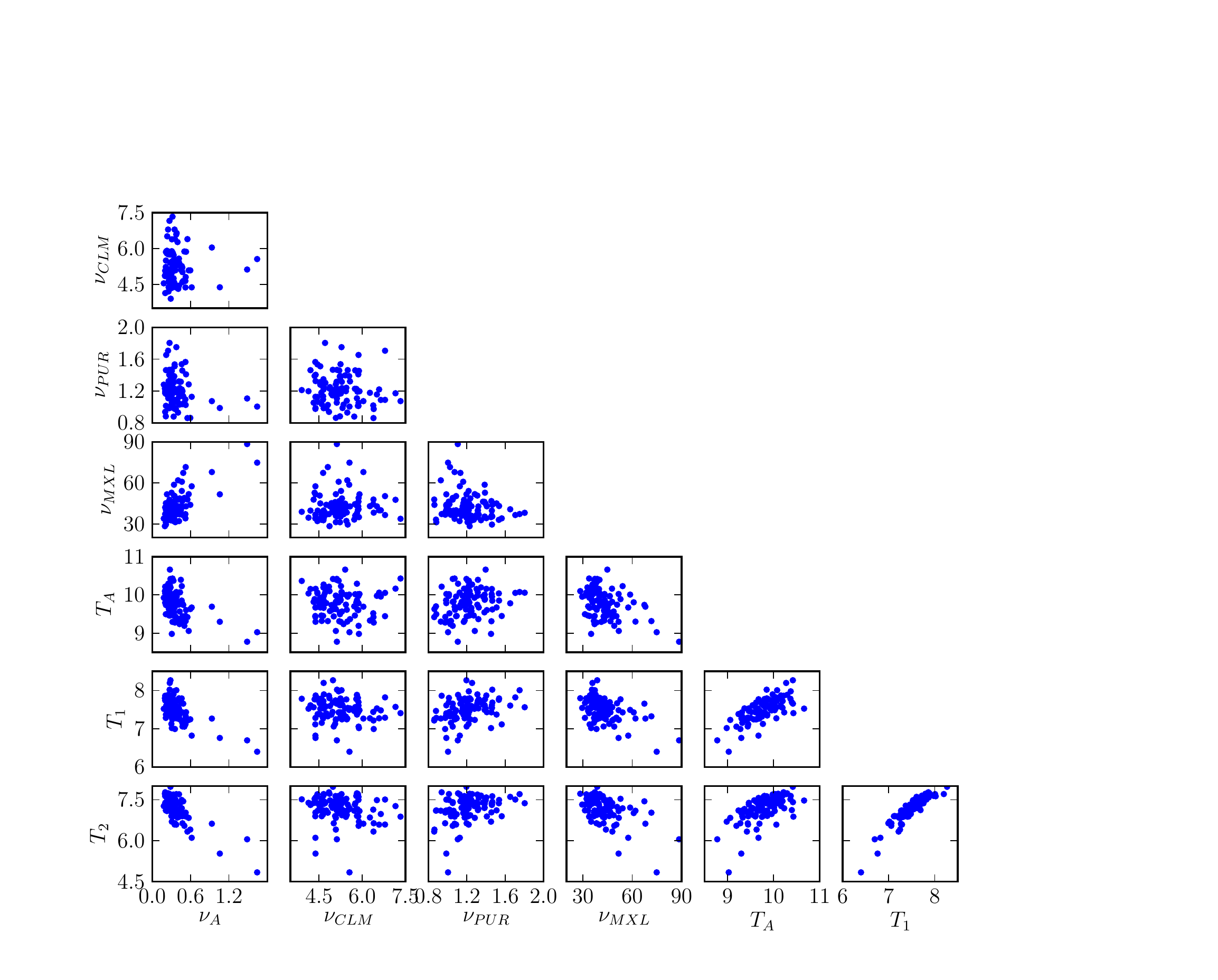}}
\caption{\label{boothists}(a) Bootstrap distributions and (b) pairwise correlations for demographic inference parameters. Vertical red bars mark the optimal parameters.}
\end{figure}

\setcounter{figure}{0}

\renewcommand{\figurename}{Text}
\begin{figure}
\caption{ Supplementary methods include additional description of statistical and filtering methods used in this article. \label{suppmet}}
\end{figure}
\renewcommand{\figurename}{Figure}
\clearpage




\section*{Text S1: Reconstructing Native American Migrations from Whole-genome and Whole-exome Data} 

\setcounter{page}{1}

\setcounter{table}{0}
\renewcommand{\thetable}{S\arabic{table}}
\subsection*{IBD and local ancestry blocks} 

For all individuals presented here, we retrieved phased local ancestry inference results generated using {\sc RFMix} as part of the 1000 Genomes project \cite{GenomesProjectConsortium:2012co}. The accuracy of the RFMix algorithm has been tested under a wide range of scenarios in \cite{Maples:2013fia}, and was found to be robust to the likely amount of divergence between the actual and proxy ancestral populations used here.  

IBD tracts across individuals were obtained using {\sc germline} \cite{Gusev:2009wp} with the `phased' (-h\_extend) option and the 1000 Genomes recombination map. Because this method tends to overcall IBD around centromeres, we discarded regions with over $100$ inferred IBD segments, and only IBD segments spanning 3Mb on each side of these regions were considered as spanning the regions. Finally, individuals NA19657, NA19661, NA19660, and NA19675 were not used in the IBD analysis due to known or inferred relatedness with other samples. 

Because the {\sc Admixture} and {\sc AS-PCA} analyses used comparisons to a Native American panel \cite{Reich:2012cs} different from the panel used to infer local ancestry in the 1000 Genomes \cite{Mao:2007gs}, we re-inferred local ancestry using the panel from  \cite{Reich:2012cs} for the purpose of these analyses.

\subsection*{{\sc Admixture} and AS-PCA}
We developed software tools to perform principal-component analysis within fractions of continental ancestry Moreno-Estrada et al, PLoS Genetics, in Press. For all {\sc admixture} and PCA analyses, we used a combination of data from the 1000 Genomes and 493 Native American samples from \cite{Reich:2012cs}. These Native American samples are drawn from 52  populations specified in supplementary Table 1 of \cite{Reich:2012cs}. We had access to genotypes at 364,470 SNPs from the Illumina 650K platform. To perform the analysis jointly with 1000 Genomes OMNI data, we restricted the analysis to the 207,430 SNPs in the intersection of the two genotyping platforms. {\sc Admixture} was run for K=2-14, and plotted following a North-South order within the NAT groups (Figure S6).


\subsection*{Calculating the site-frequency spectrum}

The original data in vcf format was downloaded from the 1000 Genomes server: \url{ftp://ftp-trace.ncbi.nih.gov/1000genomes/ftp/release/20110521}. Frequency-based demographic analyses were restricted to parts of the genome that were within the consensus exome target and that passed the strictest callability mask from the 1000 Genomes project. One individual from each pair with cryptic relatedness  were removed. This resulted in discarding 7 MXL individuals, 
NA19661,
NA19675,
NA19660,
NA19657,
NA19664,
NA19672, and
NA19726.

Diploid local ancestry assignments were obtained from the 1000 Genomes server: \url{ftp://ftp-trace.ncbi.nih.gov/1000genomes/ftp/phase1/analysis_results/ancestry_deconvolution}, for the CLM, MXL, and PUR. Diploid assignments could be any pair drawn from African, European, or Native American ancestries, or Unassigned.   

Reference and nonreference allele frequencies were tabulated for each population and ancestry, and SNPs were annotated according to Gencode version 13. Alignment with chimp was performed using {\sc liftover}, the PanTro3 chimpanzee genome, and the corresponding alignment files from UCSC website. 

To estimate the expected number of synonymous  sites in the callable region, we used the Hwang-Green mutational model \cite{Hwang:2004cf}. The Hwang-Green model provides $4^4$ mutational rates for sites with different immediate neighboring bases.  Each site was allowed to mutate proportionally to this context dependent rate, and the overall counts were normalized to obtain an average of one annotation per site. We obtained an expected total of $5,030,734$ synonymous sites.

 \subsection*{Estimating theta and the mutation rate}

When performing demographic inference using $\partial$a$\partial$i, we calculated the expected frequency spectrum assuming a unit-size initial population. All the parameters in the demographic model are scaled by the actual size of the initial population size, which can be obtained as the ratio of the number of segregating sites observed in the data, to that expected under the unit-population-size model. The first one is easily computed, but the second one requires an estimate the number of sites sequenced. From the total estimated number of $5,030,734$ synonymous sites,  we need to remove sites that would not have passed downstream filters. Because we focused on sites that had minimal amounts of Native American ancestry, we discarded a different proportion of sites for each pairwise SFS--the fraction of sites that passed all filters was $0.895$, $0.737$, and $0.693$ in the CLM-MXL, PUR-MXL, PUR-CLM  pairwise spectra. We obtain a slightly different theta value for each pairwise spectrum; we used the mean of these three values in the rest of the analysis. Finally, we need to fix either the human mutation rate, or the time of an event in our demographic model. We used 16 kya as a reference point for the population recovery into the Americas, and inferred the mutation rate based on this value.

\subsection*{Confidence intervals for demographic parameters through the SFS}

We estimated confidence intervals induced by the finiteness of the genome by bootstrapping over contiguous genomic loci. The genome was divided at every segment of at least 100kb with no sequence data.  The resulting 4493 `loci' were then sampled with replacement 100 times, and the complete inference pipeline was run on the resulting target region: estimation of the singleton error rates, inference of genetic parameters, and conversion to physical units by estimating the effective synonymous sequenced length. The resulting confidence intervals are shown on Table 1.

 \subsection*{Significance of allele frequency difference}
We calculated the $95\%$ confidence interval width $w$ for the allele frequency in the Native American components at each position in each population panel. To estimate the significance of the allele frequency difference among two groups, we fitted a Binomial distribution to the frequency and confidence interval observed in each population, and performed a parametric bootstrap with $3\times 10^7$ iterations, counting the number of occurrences where the ordering is inverted.  Using a Bonferroni correction with 29,354 tests, a bootstrap with $0$ inverted samples would yield $p<0.001.$

\subsection*{Simulating recent bottlenecks}
\label{botpur}
{\color{black}
 The piecewise-constant model of Native American population sizes considered in the main text does not allow for recent fluctuations in the Native population sizes. However, many Native American populations suffered drastic reductions in population sizes after the arrival of Europeans. We expect that the piecewise constant population sizes are effective sizes that average out the real, fluctuating population size.  If we wish to interpret these effective population sizes in terms of pre-Colombian population sizes, we must account for the possible effect of recent bottlenecks in these inference results. We also wanted to study the effect of the bottlenecks on other model parameters, such as the split times. }
 
We focused out attention on the PUR population because 1) history suggests a particularly drastic bottleneck, 2) historical data about this bottleneck is more detailed than for other populations, and 3) it has the lowest effective population size, suggesting that the recent, strict  bottleneck might explain the difference in inferred population sizes.  In Puerto Rico, the Native population prior to the arrival of Europeans was estimated at slightly above 110,000 individuals \cite{Moscoso:2008ti}. After contact, the population decreased rapidly. It is difficult to estimate the extent of this bottleneck due to incomplete and biased post-contact census data. To get an order-of-magnitude estimate, we note that the total census population of San Juan in 1673 was 1523. If we suppose that San Juan represented $10\%$ of the Island population (the figure from 1765, when such data becomes available), and that the Native American ancestry proportion in the population by then was in equal proportion to what it is today ($13\%$), we have an aggregated Native American population ($aNA$, see next section) of about 2000 Native Americans. Assuming that the effective population size bottleneck is proportional to this census population size bottleneck, we may consider a 1.5\% bottleneck lasting from 1500 to 1750. By 1765, the island population is 45,000 (aNA=$5,850$) and rapidly expands thereafter, and we consider that the bottleneck had ended by that point. 

We first attempted to estimate the size of the bottleneck from the data by re-estimating all parameters and letting the depth of the bottleneck vary. We did not find this to significantly increase the model likelihood, confirming that we do not have the power to differentiate between the no-bottleneck and the bottleneck case. To estimate the possible effect of the bottleneck on parameter inference and on the pre-Columbian population size estimates, we imposed the $1.5\%$ bottleneck and optimized all other parameters (including the population size in the PUR branch). As expected, the inferred population size in the PUR population was increased, by a factor of $3.9$. It remained lower than the CLM estimated population size (without bottleneck). The maximum change in any other parameter was a $10\%$ increase in the PUR/CLM split time.

 To further test the robustness of split time inference to the presence of bottlenecks, we introduced a second bottleneck of identical duration and depth in the CLM population. We found a similar 4-fold increase of CLM and PUR pre-bottleneck population sizes, relative to the no-bottleneck models. Both population size estimates remained well below the MXL estimate. We found modest parameter changes for other parameters: a $20\%$ increase in the MXL population size estimate, and increases of $9.9\%$ in the MXL split time and $ 8.4\%$ in the CLM/PUR split time.  
   
Finally, we can obtain an order-of-magnitude estimate of the change in effective population size due to a bottleneck using the harmonic mean formula for drift: if the bottleneck lasts a fraction $\rho$ of the current time period, and the population reduces to size $\alpha N$, we get 
$\hat N_{e}=\frac{N}{\rho/\alpha +(1-\rho)}$. In the current model, $\rho=0.019$, and $\alpha=0.015$, so that $\hat N_e=2.2 N$. However, this estimate does not account for new mutations occurring during the time interval. Because such mutations represent a considerable proportion of mutations in our model, the harmonic mean estimates don't provide accurate results, but may help in quickly assessing the effects of different bottleneck models.  

\subsection*{Aggregated effective population size}
\label{aggreg}
We wish to model the allele frequency within the Native American component of admixed populations as if it was evolving under a randomly mating population of a given effective size. The natural choice would be to use an effective population size equal to the average number of Native American haplotypes in the population. We call this the aggregate Native American population size, since it represents an effective size across of Native American ancestry aggregated over all individuals.  In this section, we show that this is reasonable.

Given that we have $N$ Native American alleles at generation $t$, $pN$ of which carry allele $a$ and $(1-p)N$ of which carry allele $A,$ the variance in $p'$, the proportion of allele $a$ in the next generation, can be expressed using the law of total variance $\var(p')=  E_N'(\var(p'|N')) + \var_N' (E(p'|N')) $, where $N'$ is the number of Native American alleles at the next generation. Because  $E(p'|N')=p$ is independent of $N'$, the second term is zero. 
The first term is  $$E_N'(\var(p'|N'))=E_N'(p (1-p)/N')= p (1-p) E_N'(1/N').$$  The latter term is infinite because of the ever so slight probability that no Native haplotype remains, which would lead to an indeterminate value for p. Because we do not calculate allele frequencies in such cases, we find that to a very good approximation (for $N>100$) $E_N'(1/N')\simeq 1/N$.  Thus we find $\var(p')=p(1-p)/N.$ As could be expected, the variation in allele frequency behaves roughly as it would if we tracked allele frequencies in a total population size of size $N$, the expected number of Native American alleles per site. Because the proportion of Native American ancestry can vary from site to site due to drift, we have a drift term for $\var(p')$. In other words, there is drift on the drift parameter. However, given the short time since admixture in the present study, we will neglect such second-order drift.

\subsection*{Significance of IBD vs Ancestry results}
 
We wish to determine whether long IBD tracts have a higher density of ancestry switch-point, given that they are likely to have more recent TMRCAs. To do so, we first discarded 2cM from each edge of IBD tracts, because the IBD boundaries are likely to also be ancestry switch-points, and small errors in the estimated positions of switch-points can inflate the number of inferred switches within the tracts. We calculated the number and density of switch-points in the interior region. We then sorted the IBD segments according to length, grouped them in blocks of $n$ IBD segments, and computed the mean length and switch-point density. This grouping allows us to calculate the uncertainty in each block via the bootstrap--we chose $n=40$ in the MXL and $n=200$ in the other two populations. Much smaller block sizes result in risking some bins containing only IBD segments without switch points, making the bootstrap analysis meaningless. Because MXL has only $307$ IBD segments of sufficient length, smaller block sizes had to be used. 
  
Once the bin-specific variances have been estimated, we computed a linear regression on the mean values for each bin using the original sample and 1000 bootstrapped samples. The reported p-values are the fraction of bootstrap instances that have non-positive correlation. 

 Results appeared robust to increasing $n$ in populations where a signal was observed, PUR ($n=400$, $p<0.001$) and MXL ($n=153$, $p=0.012$). CLM remained insignificant.  
 
 \subsection*{Confidence intervals and goodness-of-fit in {\sc tracts}}
 Reference \cite{Gravel:2012ip} described a likelihood-ratio test to compare different migration models, but not a goodness-of-fit test. In our model assumptions, the number of tracts $t_i$ in each length bin is a Poisson variable with mean $e_i$ given by the model expectation. Thus we can calculate Pearson's $\chi^2$ statistic 
 \begin{equation}
 X=\sum_{i=1}^B \frac{(t_i-e_i)^2}{e_i},
 \end{equation}
 where $B$ is the total number of bins after  bins with less than 10 expected counts per bin in each population have been pooled. We model this as a $\chi^2_{B-1-n}$ distribution, with $n$ the number of fitted parameters.

As a starting point in CLM and PUR, we considered a model in which Europeans and Native Americans first form a panmictic admixed population, which subsequently receives migrants from an African source population. {\color{black} The timing and magnitude of each migration is chosen to maximize the model likelihood. This model has four free parameters: timing and ancestry proportions at the first generation, and timing and magnitude of the African migration. }

 In Puerto Ricans, we found that the model is significantly improved upon if a second pulse of migration is allowed for both European and African ancestry: adding a pulse of African migrants at the population onset improves the log-likelihood by 17 units, and the second European migration epoch further improves the log-likelihood by 13 units. This more than justifies the three extra parameters according to the Bayesian information criterion with 150 data points. The p-value of a $\chi^2$ goodness-of-fit test for the final optimized model, displayed in Figure 3a, was 0.50, indicating that this model accurately describes the data. 

In the CLM, incorporating additional recent migration from both Europeans and Native Americans improves the fit: the additional European pulse adds 53 log-likelihood units, whereas the Native American pulse adds 14, again justifying the most complex model using the Bayesian information criterion. The final $\chi^2$ p-value is $0.47$. The best-fitting model is displayed in Figure 3b.   

In MXL, the $\chi^2$ goodness-of-fit $p$-value was $0.017$ for the model considered in \cite{Kidd:2012fi}, indicating that it may be possible to marginally improve upon this model.

{\color{black} To estimate confidence intervals on specific parameters of the migration model, we performed a bootstrap analysis by sampling \emph{individuals} with replacement. Thus these confidence intervals are robust to population structure. The optimal parameters and confidence intervals are provided in Figure \ref{figsuppCI} }

\begin{table*}
\centering
\color{black}
\begin{tabular}{ccc|ccc|ccc}
&PUR&&&CLM&&&MXL&\\
param.&estim.&$95\%$ CI & param. &estim.&$95\%$ CI&param. &estim.&$95\%$ CI  \\
\hline
$T_0$& $14.9$& $(14.2, 15.9) $&$T_0$& $13.0$& $(12.5, 13.9) $&$T_0$& $15.1$& $(13.7, 17.1) $\\
$p_{Af0}$& $0.103$& $(0.085, 0.132) $&$p_{Eu0}$& $0.690$& $(0.659, 0.712) $&$p_{Af0}$& $0.109$& $(0.078, 0.160) $\\
$p_{Eu0}$& $0.702$& $(0.606, 0.735) $&$p_{Na0}$& $0.310$& $(0.288, 0.341) $&$p_{Eu0}$& $0.453$& $(0.405, 0.496) $\\
$p_{Na0}$& $0.195$& $(0.169, 0.261) $&$T_1$& $9.6$& $(8.7, 10.8) $&$p_{Na0}$& $0.438$& $(0.385, 0.480) $\\
$T_1$& $6.8$& $(5.9, 8.8) $&$p_{Af1}$& $0.077$& $(0.056, 0.107) $&$p_{Eu}$& $0.037$& $(0.028, 0.046) $\\
$p_{Af1}$& $0.042$& $(0.017, 0.066) $&$T_2$& $4.8$& $(4.0, 6.0) $&$p_{Na}$& $0.036$& $(0.027, 0.045) $\\
$p_{Eu1}$& $0.268$& $(0.150, 0.445) $&$p_{Eu2}$& $0.141$& $(0.098, 0.213) $&&&\\
&&&$p_{Na2}$& $0.013$& $(0.003, 0.035) $&&&\\
\end{tabular}
\caption{Parameters and confidence intervals for migration parameters inferred using {\sc tracts}. \label{figsuppCI} $T$'s are time in generations and $p$'s are proportion of migrants per generation. $p$'s with numbered labels correspond to punctual migrations at the correspondingly numbered time. Because the models allow only migrations at integer times, migrations occurring at non-integer times were distributed over neighboring integer migrations with higher weight given to the more nearby value. $p$'s with no numbered labels (in MXL) correspond to continuous migrations. Migration proportions at $T_0$ are constrained to sum to one. In MXL, an additional constraint was $p_{Eu0}/p_{Na0}=p_{Eu}/p_{Na}.$  }
\end{table*}

   \subsection*{Timing recombinations and TMRCA using IBD and ancestry patterns}
 Formally,  the likelihood function for the demographic model $\theta$ and TMRCA $t$, given an IBD length of $\ell$ and an ancestry pattern $a$ is $$P(a,\ell\| t, \Theta)=P(a\| \ell, t, \Theta)P(\ell \| t, \Theta).$$ The first expression can be obtained using a Markov model developed in \cite{Gravel:2012ip}, the second using the IBD models developed in \cite{Palamara:2012LD}.  Because the amount of IBD increases rapidly with sample size, we expect such approaches to be particularly suited for large genotyping cohorts.

\end{document}